\documentclass[11pt]{article}
\usepackage{amsmath}
\usepackage{amsfonts}
\usepackage{amsxtra}
\usepackage{verbatim}
\usepackage{amssymb}
\usepackage{amsthm}
\usepackage{fullpage}
\usepackage{epsfig}
\usepackage{calc}
\usepackage{multirow}
\usepackage{bbm}
\usepackage{palatino}
\usepackage{float}
\usepackage{mathtools}
\usepackage{stmaryrd}
\usepackage{euscript}
\usepackage{paralist}
\usepackage[normalem]{ulem}
\usepackage[margin=1in]{geometry}

\usepackage[%
                  pagebackref,
                  letterpaper=true,
                  colorlinks=true,
                  pdfpagemode=none,
                  urlcolor=blue,
                  linkcolor=blue,
                  citecolor=blue,
                  pdfstartview=FitH]
                  {hyperref}

\usepackage{cleveref}

\def\labs{[k]}

\makeatletter 
\global\let\@myarticle\@true
\makeatletter
\renewcommand{\paragraph}{%
  \@startsection{paragraph}{4}%
  {\z@}{0.0ex \@plus 1ex \@minus .2ex}{-0.25em}%
  {\normalfont\normalsize\bfseries}%
}
\makeatother
\def\SDPOPT{\mathrm{OPT}_{\mathrm{SDP}}}

\def\citeyear{\cite}

\def\triangleq{\overset{\mathrm{def}}{=}}

\newcommand{\myparagraph}[1]{\vspace{2ex} \paragraph{#1}}

\let\eps=\varepsilon

\ifx\@myarticle\@true
	\newtheorem{theorem}{Theorem}[chapter]	
\else
	\newtheorem{theorem}{Theorem}[section]	
\fi

\newtheorem{claim}[theorem]{Claim}
\newtheorem{corollary}[theorem]{Corollary}
\newtheorem{definition}[theorem]{Definition}

\newtheorem{lemma}[theorem]{Lemma}
\newtheorem{notation}[theorem]{Notation}

\newtheorem{observation}[theorem]{Observation}
\newtheorem{proposition}[theorem]{Proposition}

\theoremstyle{definition}
\newtheorem{remark}[theorem]{Remark}

\crefname{assumption}{Assumption}{Assumptions}
\crefname{conjecture}{Conjecture}{Conjectures}
\crefname{claim}{Claim}{Claims}
\crefname{corollary}{Corollary}{Corollaries}
\crefname{definition}{Definition}{Definitions}
\crefname{example}{Example}{Examples}
\crefname{exercise}{Exercise}{Exercises}
\crefname{fact}{Fact}{Facts}
\crefname{lemma}{Lemma}{Lemmas}
\crefname{notation}{Notation}{Notations}
\crefname{note}{Notes}{Notes}
\crefname{observation}{Observation}{Observations}
\crefname{proposition}{Proposition}{Propositions}
\crefname{problem}{Problem}{Problems}
\crefname{question}{Question}{Questions}
\crefname{remark}{Remark}{Remarks}

\DeclareMathOperator{\poly}{poly}

\newcommand{\R}{\mathbb{R}}

\DeclareMathOperator{\nll}{null}
\newcommand{\symm}{\mathbb{S}}
\newcommand{\sympm}{\mathbb{S}_{+}}

\newcommand{\dims}{\Upsilon}

\let\proj=\Pi

\DeclareMathOperator{\tr}{Tr}
\newcommand{\unit}[1]{\overline{#1}}
\newcommand{\dunit}[1]{\overline{\overline{#1}}}
\DeclareMathOperator{\diag}{diag}
\DeclareMathOperator{\spn}{span}
\def\Lnorm{\mathcal{L}}
\def\Anorm{\mathcal{A}}
\newcommand{\ind}[1]{\mathbbm{1}_{#1}}

\DeclareMathOperator{\rank}{rank}
\let\es=\emptyset
\newcommand{\bestSi}{{\mathcal{S}^\ast}}
\newcommand{\bestS}{{\mathcal{S}^\ast}}
\newcommand{\lasserrei}[3]{\mathrm{Moment}_{#1}(#2,#3)}
\newcommand{\bestSfi}[1]{\|\xvec_{\circ| \bestS(\circ)}\|^2}
\newcommand{\xint}{\mathbf{x}} %
\newcommand{\randx}{{\scriptstyle \mathcal{X}}}
\newcommand{\xvec}{\mathrm{X}}%
\newcommand{\yvec}{\mathrm{Y}}%

\newcommand{\wvec}{\mathrm{W}}%
\newcommand{\xmat}{\mathrm{X}}%
\newcommand{\ymat}{\mathrm{Y}}%
\newcommand{\zmat}{\mathrm{Z}}%
\newcommand{\wmat}{\mathrm{W}}%

\def\opt{\mathsf{OPT}}
\DeclareMathOperator{\OPT}{OPT}%

\newcommand{\nusc}{\textsc{\sf Non-Uniform Sparsest Cut}}

\newcommand{\lh}{\textrm{Lasserre Hierachy}}

\DeclareMathOperator*{\st}{st}
\DeclareMathOperator*{\argmin}{argmin}

\newcommand{\probs}[2]{\mathrm{Prob}_{#1}\big[ #2 \big]}
\newcommand{\expcts}[2]{\mathbb{E}_{#1}\big[ #2 \big]}
\newcommand{\vars}[2]{\mathrm{Var}_{#1}\big[ #2 \big]}

\newcounter{alg-count}
\floatstyle{ruled}
\newfloat{program}{thp}{lop}%
\floatname{program}{Algorithm}

\crefname{alg-count}{step}{steps}
\crefname{program}{Algorithm}{Algorithms}

\makeatother

\title{Rounding Lasserre SDPs using column selection and
  spectrum-based approximation schemes for graph partitioning and
  Quadratic IPs
} 
\author{Venkatesan Guruswami 
	\thanks{The research of V. Guruswami was supported in part by NSF
     CCF-1115525 and a Packard Fellowship.}
	\\ 
	\url{guruswami@cmu.edu}\\
	Computer Science Department, \\ 
	Carnegie Mellon University, \\ Pittsburgh, PA.
	\and
	Ali Kemal Sinop
	\thanks{The research of A. K. Sinop
     was supported in part by NSF DMS-1128155, 
     NSF CCF-1115525 and the MSR-CMU Center for Computational
     Thinking. Part of this work was
     done when A. K. Sinop was at Carnegie Mellon University. }
	 \\
	\url{asinop@cs.cmu.edu}
	\\
	School of Mathematics, \\
	Institute for Advanced Study, \\
	Princeton, NJ.
	 }
\begin{document}
\maketitle
\thispagestyle{empty}
\begin{abstract}
  We present an approximation scheme for minimizing certain Quadratic
  Integer Programming problems with positive semidefinite objective
  functions and global linear constraints. This framework includes
  well known graph problems such as Minimum graph bisection, Edge
  expansion, Sparsest Cut, and Small Set expansion, as well as
  the Unique Games problem. These problems are notorious for the
  existence of huge gaps between the known algorithmic results and
  NP-hardness results.  Our algorithm is based on rounding
  semidefinite programs from the Lasserre hierarchy, and the analysis
  uses bounds for low-rank approximations of a matrix in Frobenius
  norm using columns of the matrix.

  For all the above graph problems, we give an algorithm running in
  time $n^{O(r/\eps^2)}$ with approximation ratio
  $\frac{1+\eps}{\min\{1,\lambda_r\}}$, where $\lambda_r$ is the
  $r$'th smallest eigenvalue of the normalized graph Laplacian
  $\Lnorm$. In the case of graph bisection and small set expansion,
  the number of vertices in the cut is within lower-order terms of the
  stipulated bound.  Our results imply $(1+O(\eps))$ factor
  approximation in time $n^{O(r^\ast/\eps^2)}$ where is the number of
  eigenvalues of $\Lnorm$ smaller than $1-\eps$ (for variants of
  sparsest cut, $\lambda_{r^\ast} \ge \OPT/\epsilon$ also suffices,
  and as $\OPT$ is usually $o(1)$ on interesting instances of these
  problems, this requirement on $r^\ast$ is typically weaker). This
  perhaps gives some indication as to why even showing mere
  APX-hardness for these problems has been elusive, since the
  reduction must produce graphs with a slowly growing spectrum (and
  classes like planar graphs which are known to have such a spectral
  property often admit good algorithms owing to their nice structure).
  
  For Unique Games, we give a factor $(1+\frac{2+\eps}{\lambda_r})$
  approximation for minimizing the number of unsatisfied constraints
  in $n^{O(r/\eps)}$ time, improving upon an earlier bound for solving
  Unique Games on expanders. We also give an algorithm for independent
  sets in graphs that performs well when the Laplacian does not have
  too many eigenvalues bigger than $1+o(1)$.
\end{abstract}

\newpage

\section{Introduction}
\label{sec:intro}

\maketitle
\thispagestyle{empty}

\newpage

\section{Introduction}
\label{sec:intro}

\section{Preliminaries}
\label{sec:prelim}
We now formally define the notation and terminology that will be
useful to us in the paper.

\myparagraph{Sets} %
Given set $A$ and positive integer $k$, we use $\binom{A}{k}$
(resp. $\binom{A}{\le k}$) to denote the set of all possible size $k$
(resp. size at most $k$) subsets of $A$.  We use $\R_+$ to denote the
set of non-negative reals.

\myparagraph{Euclidean Space} Given row set $B$, we use $\R^B$ to
denote the set of real vectors where each row (axis) is associated
with an element of $B$.
For any vector $\xvec \in \R^B$, its coordinate at axis $b \in B$ is
denoted by $\xvec(b)$.  Let $\|\xvec\|_p$ be its $p^{th}$ norm with
$\|\xvec\|\triangleq \|\xvec\|_2$, and $\xvec^T$ be its transpose.
Finally for any $\xvec, \yvec\in \R^B$, let $\langle \xvec,
\yvec\rangle = \xvec^T \yvec$ be their inner product $\sum_{b\in B}
\xvec_b \yvec_b$.
 
\myparagraph{Matrices} Given %
row set $R$ and column set $C$, we use $\R^{R,C}$ \footnote{We chose
  this notation over the conventional one ($\R^{R\times C}$) so as to
  prevent ambiguity when the rows, $R$, or columns, $C$, are Cartesian
  products themselves.} to denote the set of real matrices whose
rows and columns are associated with elements of $R$ and $C$,
respectively.  Given matrix $X\in \R^{R, C}$, for any $r\in R,
c\in C$, we will use $X_{r,c}\in \R$ to denote entry of $X$ at
row $r$ and column $c$. For convenience, we use $X_c\in \R^{R}$ to
denote the vector corresponding to the column $c$ of $X$. Likewise
given subset of columns of $X$, $S\subseteq C$, we use $X_S\in
\R^{R, S}$ to denote the matrix corresponding to the columns $S$ of
$X$.  Given matrix $X$, we use $\|X\|_F$, $\tr(X)$ and
$X^T$ to denote Frobenius norm of $X$, its trace and transpose.
Finally we use $\xmat^{\proj}$ and $\xmat^{\perp}$ to denote the
projection matrices onto the column span of $\xmat$, $\spn(\xmat)$,
and its orthogonal complement. For a matrix $A$, we denote by $\nll(A)$ its right nullspace, i.e., the set of vectors $x$ for which $Ax=0$.

We will use $\symm^C$ and $\sympm^C$ to denote the set of symmetric
and positive semidefinite matrices, respectively.
%
\myparagraph{Matrix Width} For any matrix $A \in \R^{B,C}$, we will
refer to the maximum number of non-zero entries among rows of $A$ as
the {\bf width} of $A$.

\myparagraph{Positive Semidefinite (PSD) Ordering} Given a symmetric
matrix $\xmat\in\symm^A$, we say $\xmat$ is a PSD matrix (i.e. $\xmat
\in \sympm^A$), denoted
by $\xmat \succeq 0$, iff $\yvec^T \xmat \yvec \ge 0$ for all $\yvec
\in \R^{A}$. 

\begin{remark}[Convenient matrix notation]
  One common expression we will use throughout this paper is the
  following. For matrices $\xmat=[\xvec_u] \in \R^{\dims \times R_0}$ and $M\in
  \R^{R_1\times R_1}$ with $R_1 \subseteq R_0$\footnote{If $M$
    corresponds to a proper principal minor of $\xmat^T \xmat$, we assume $M$
    is padded with enough $0$'s before multiplication.}:
  \[ \tr( \xmat^T \xmat M) = \sum_{\substack{u\in R_0\\ v\in R_0 }}
    M_{u,v} \langle \xvec_u, \xvec_v\rangle \ . \]
  \noindent Note that if $M$ is positive semidefinite, i.e. $M\succeq
  0$, then $\tr(\xmat^T \xmat M) \ge 0$.\footnote{The use of this inequality
    in various places is the reason why our analysis only works for
    minimizing PSD quadratic forms.} {\hfill $\Box$}
\end{remark}

\myparagraph{Eigenvalues} Given symmetric matrix $M \in \symm^{A}$,
for any integer $i \le |A|$, we define its $i^{th}$ smallest and
largest eigenvalues of $M$ as the following, respectively:
\begin{align}
\lambda_i(M ) \triangleq & \max_{\rank(\zmat) \le i-1} \min_{\wvec\perp \zmat, 
\wvec \neq 0}
\frac{ \wvec^T M \wvec }{\wvec^T \wvec}, \nonumber
\\
\sigma_i(M ) \triangleq& \min_{\rank(\zmat) \le i-1} \max_{\wvec\perp \zmat, 
\wvec \neq 0}
\frac{ \wvec^T M \wvec }{\wvec^T \wvec}. \label{eq:sigma-i-def}
\end{align}
\myparagraph{Generalized Eigenvalues} Given $L\in\symm^A$ and $M \in
\sympm^{A}$ with $\nll(M) \subseteq \nll(L)$, for any integer $i \le
\rank(M)$, we define $i^{th}$ smallest generalized eigenvalue of $L$
and $M$ as the following:
\begin{equation} %
\label{eq:glambda-i-def}
  \lambda_i(L,M) \triangleq
    \max_{\rank(\zmat) \le i-1} \min_{M \wvec\perp \zmat, 
      M \wvec \neq 0}
    \frac{ \wvec^T L  \wvec }{\wvec^T M \wvec}
\end{equation} 
Observe that, for $r$ being the nullity of $M$, $\lambda_i(L,M) =
\lambda_{i+r}((M^{\dagger})^{1/2}\  L  \ (M^{\dagger})^{1/2})$.
\myparagraph{Graphs} We assume all graphs are simple, undirected and
edge-weighted with non-negative weights. We associate each graph
$G=(V,C)$ with its edge weight function of the form $C:\binom{V}{2} \to
\R_+$, where we use $C_{u,v}$ to denote the weight of edge between $u$
and $v$.
\myparagraph{Adjacency, Laplacian and Incidence Matrices} Given an edge-weighted graph
$G=(V,C)$ with no self loops, we define its adjacency matrix, $A_G \in \symm^{V}$; degree matrix $D_G$;
Laplacian matrix, $L_G\in \symm^{V}$; and edge-node incidence matrix,
$\Gamma_G \in \R^{\binom{V}{2}, V}$, as: %
\[
(A_G)_{u,v} = C_{u,v}.
\] 
\[
(D_G)_{u,v} = \begin{dcases*}
		\sum_{w} C_{u,w} & if $u=v$, \\
		0 & else
	\end{dcases*} \]
\[ L_G = D_G - A_G, \quad \text{and} \]
\[
(\Gamma_G)_{uv,w} = \begin{dcases*} \sqrt{C_{uv}} & if $w =
  \min(u,v)$, \\
  -\sqrt{C_{uv}} & if $w =
  \max(u,v)$, \\
  0 & else.
\end{dcases*}
\] 
Observe that:
\begin{inparaenum}[(i)]
\item $\Gamma_G$ has width-$2$.
\item $L_G = \Gamma_G^T \Gamma_G$, hence $L_G \in \sympm^V$. 
\item For any $X =
[X_u] \in \R^{\Upsilon, V}$, $\tr\left[ X^T X
  L_G\right] = \tr\left[ X^T X \Gamma_G^T \Gamma_G\right] = \|X \Gamma_G^T\|^2_F
= \sum_{u<v} C_{u,v} \left\| X_u-X_v \right\|^2$.
\item $\lambda_1(L_G) = 0$. 
\item If $G$ is connected, then $\lambda_2(L_G) > 0$.
\end{inparaenum}

The normalized Lapalcian matrix $\Lnorm_G$ is defined as $\Lnorm_G = D^{-1/2} L_G D^{-1/2}$. We will often omit the subscript $G$ from these matrices as the graph will be clear from the context.

%
%
%
%
%
%
%
%
%
%
%
%
%
%
%
%
%
%
%
%
%
%
%
%
%
%
%

%
%

%
%
%
%
%
%
%
%
%
%
%
%
%
%
%
%
%
%
%
%
%
%
%
%
%
%
%
%
%
%
%
%
%
%
\subsection{Lasserre Hierarchy of Semidefinite Programming
  Relaxations}
\label{app:conditioning-tools}
\label{sec:lasserre-defn}
We present the formal definitions of the family of SDP relaxations,
tailored to the setting of the problems we are interested in, where
the goal is to assign to each vertex/variable of a set $V$ a label
from $[k]=\{0,1,\dots,k-1\}$. One can see that this relaxation is
equivalent to the SDP hierarchy as given in~\cite{las02} by basic
inclusion-exclusion (for a formal proof, see the decomposition theorem
in~\cite{kmn10}).

\begin{definition}[SDP Relaxation]
  \label{def:las-sdp}
  Given a set of variables $V$, a set $[k]=\{0,1,2,\ldots,k\}$ of
  labels and an integer $r \ge 0$, $\xmat = (\xvec_S(f)\in
  \R^{\Upsilon})$ is said to satisfy $r^{th}$ moment constraints on
  $k$ labels, denoted by \[ \xmat \in \lasserrei{r}{V}{k} \ ,
  \] %
  if it satisfies the following conditions:
  \begin{enumerate}
    \itemsep=0.7ex
  \item For each set $S\in\binom{V}{\le r+1}$ and $f:S\to \labs$,
    there exists a function $\xvec_S:[k]^{S}\to \R^{\Upsilon}$ that
    associates a vector of some finite dimension $\Upsilon$ with each
    possible labeling of $S$.  We use $\xvec_S(f)$ to denote the
    vector associated with the labeling $f\in [k]^S$.  For singletons
    $u\in V$, we will use $\xvec_u(i)$ and $\xvec_u(i^u)$ for $i\in
    [k]$ interchangeably.
    Similarly, when $k=2$, we will use $\xvec_u$ instead of $\xvec_u(1)$.

    For $f \in [k]^S$ and $v\in S$, we use $f(v)$ as the label $v$
    receives from $f$.  Also given sets $S$ with labeling $f\in [k]^S$
    and $T$ with labeling $g\in[k]^T$ such that $f$ and $g$ agree on
    $S\cap T$, we use $f\circ g$ to denote the labeling of $S\cup T$
    consistent with $f$ and $g$: If $u\in S$, $(f\circ g)(u) = f(u)$
    and vice versa.
  \item Let $\labs^{\es} = \{\top\}$ where $\top$ denotes the (only)
    labeling of empty set with $\xvec_{\es}(\top)=\xvec_{\es}$.
  \item $\langle \xvec_S(f), \xvec_T(g)\rangle = 0$ if there exists
    $u\in S\cap T$ such that $f(u)\neq g(u)$.
  \item \label{def:las-sdp:consistent} $\langle \xvec_S(f),
    \xvec_T(g)\rangle = \langle \xvec_A(f'), \xvec_B(g')\rangle$ if
    $S\cup T=A\cup B$ and $f\circ g = f'\circ g'$.
  \item For any $u\in V$, $\sum_{j\in [k]} \|x_u(j)\|^2 = \|x_\es\|^2$. 
  \item (implied by above constraints) For any $S\in\binom{V}{\le r+1}$,
    $u\in S$ and $f\in [k]^{S\setminus\{u\}}$, $\sum_{g\in [k]^u}
    \xvec_{S}(f\circ g) = \xvec_{S\setminus\{u\}} (f)$.
  \end{enumerate} 
\end{definition}
It was shown in~\cite{las02} that one can handle any polynomial
constraints in the following way:
\begin{definition}[Inequality Constraints]
  \label{def:linear-ineq-constraints}
  Given a degree-$d$ polynomial constraint of the form
  \[
  z(\xint) = 
  \sum_{S\in \binom{V}{\le d}, f:S\to \labs} z_f 
  \prod_{u\in S} \xint_u(f(u)) \ge 0,
  \] we say $\xmat \in \lasserrei{r}{V}{k}$ satisfies $z$:
  \[
  \xmat\ast z \in \lasserrei{r}{V}{k},
  \]
  if there
  exists $\ymat=[\yvec_S(f)] \in \lasserrei{r-d}{V}{k}$ such that 
  \[
  \|\yvec_T(g)\|^2 = \sum_{S\in \binom{V}{\le d}, f:S\to\labs} z_S
  \|\xvec_{S\cup T}(f\circ g)\|^2
  \] for any $g:T\to\labs$ with $|T|\le r-d$.
\end{definition}
In the case of equality constraints, there is a simpler way to enforce
which is also equivalent:
\begin{definition}[Equality Constraints]
  \label{def:linear-eq-constraints}
  Given a degree-$d$ polynomial constraint of the form
  $z(\xint) = 0$,
  we say $\xmat \in \lasserrei{r}{V}{k}$ satisfies $z$ if 
  \[
  \xmat \cdot z = 
  \sum_{f:S\to\labs} z_f \xvec_S(f) = 0.
  \] 
\end{definition}
\begin{claim}
  \label{thm:ineq-to-eq}
  Given a degree-$d$ polynomial $z$, 
  for any $\xmat \in \lasserrei{r}{V}{k}$,
  $\{\xmat \ast z, \xmat \ast -z\} \subset \lasserrei{r-d}{V}{k}$ 
  if and only if 
  $\xmat \cdot z = 0$.
\end{claim}
\begin{definition}[Conditioning]
  \label{def:cond} Given $S\subseteq [n]$ and $f\in \labs^S$ with
  $\xvec_S(f)\neq 0$, we define the vectors conditioned on $f$ as
  follows. For any $T\subseteq [n]$ and $g\in \labs^T$, the vector
  $\xvec_{T\mid f}(g)$ is given by:
  \[
  \xvec_{T\mid f}(g) \triangleq \frac{\xvec_{S\cup T}(f\circ g)}{\|
    \xvec_S(f)\|}.
  \] 
  We will use $\xmat_{\mid f} = [\xvec_{T\mid f}(g)]$ to denote the
  corresponding matrix.
\end{definition}
Formally the conditional vectors $\xvec_{T|f}(g)$ correspond to
relaxations of respective indicator variables. Thus such vectors
behave exactly in the same way with non-conditional vectors. Some of
these properties are given in the following easy claim, whose proof we
skip.  
\begin{proposition} \label{clm:cond-vecs} Given $\xmat \in
  \lasserrei{r}{V}{k}$, for any $f\in \labs^S$ with $\xvec_S(f) \neq
  0$, the following are true:
  \begin{enumerate}[(a)]
  \item \label{thm:cond:0} $\xmat_{\mid f} \in \lasserrei{r-|S|}{V}{k}$.
  \item \label{thm:cond:1} If $\xmat \ast z \in \lasserrei{r-d}{V}{k}$
    for some degree-$d$ polynomial $z$, then $\xmat_{\mid f} \ast z
    \in \lasserrei{r-|S|-d}{V}{k}$.
  \item \label{thm:cond:2} For any $g \in \labs^T$,
    $
    \left(\xmat_{\mid f}\right)_{\mid g} = \xmat_{\mid f\circ g}.
    $
  \end{enumerate}
\end{proposition}
Assume that some labeling $f_0 \in \labs^{S_0}$ to $S_0$ has been
fixed, and we further sample a labeling $f$ to $S$ with probability
$\|\xvec_{S|f_0}(f)\|^2$ (i.e., from the conditional probability
distribution of labelings to $S$ given labeling $f_0$ to $S_0$). The
following defines a projection matrix which captures the effect of
further conditioning according to the labeling to $S$. For a nonzero
vector $v$, we denote by $\unit{v}$ the unit vector in the direction
of $v$.

\begin{notation}%
  \label{not:pis}
  Given $f_0 \in \labs^{S_0}$ and $S\subseteq[n]$, let 
  \[
  \Pi_{S\mid f_0} \triangleq \sum_{f:\xvec_{S|f_0}(f) \neq 0}
  \unit{\xvec_{S\mid f_0}(f)} ~ \cdot ~ \unit{\xvec_{S\mid f_0}(f)}^T.
  \]
  Similarly let $\Pi_S^{\perp} \triangleq I - \Pi_S$ where $I$ is the
  identity matrix of the appropriate dimension.
\end{notation}

\subsection{Overview of Our Rounding}
\label{sec:rounding-overview}
Consider any solution satisfying $(r+2)^{th}$ moment constraints. For any
set of $r$ variables, $S$, SDP solution gives us a distribution over
$f:S\to \labs$. Moreover, conditioned on having labelled $S$ according
to $f$, we know that the conditional vectors as in \Cref{def:cond}
satisfies $2^{nd}$ moment constraints. %

This suggests a natural approach for rounding: 
\begin{enumerate}
\itemsep=0.5ex
\item Fix some set of ``seeds'', $S$.
\item Choose a labeling for $S$, $f:S\to \labs$ with probability
  $\|\xvec_S(f)\|^2$.
\item For every node $u\in V$, choose a label such that the marginal
  probability of assigning label $i$ to $u$ is $\|\xvec_{u\mid
    f}(i)\|^2$.
\end{enumerate}

A simple way to choose labelings so as to satisfy the marginal
probability condition is by sampling labels independently at random
with respective probabilities. This is not the only way for choosing
labels, and indeed our rounding for \nusc\ problem uses a different
rounding scheme.

For graph partitioning type problems, we can upper bound the
probability of not satisfying an edge constraint with the sum of SDP
value and the (co-)variance of associated variables
(see~\Cref{app:conditioning-tools} for formal proof). 

Our main observation is that, the expected variance can be related to
a highly geometric quantity involving only the ``singleton'' vectors:
The expected variance on $\xint_u(i)$ is upper bounded by the squared
distance of $\xvec_u(i)$ to the span of vectors in the seed set.

In particular, the total variance is now upper bounded by how well the
vectors in seed set approximate the singleton matrix in Frobenius
norm. Using the column selection methods from
from~\cite{gs11-svd,bdm11}, we can finally upper bound the total
variance of labeling from such distribution by the sum of largest $r$
eigenvalues of associated Gram matrix. Finally by using trace
inequalities, we can lower bound this sum in terms of associated
constraint graph's spectra.

\subsection{Faithful Rounding}
\label{sec:faithf-rand-var}
All our rounding algorithms will follow the same outline. We will
assume that we are given a subset of {\bf seeds}, $S\subset V$, with
corresponding vectors $(\xvec_S(f)\mid f:S\to\labs)$ satisfying some
$r$ rounds of \lh\ constraints on labels $\labs$. We proceed by
choosing a seed labelling, $f: S\to \labs$, and then {\bf propagating}
$f$ to other nodes in such a way that the probability of some $u\in V$
receiving label $j\in\labs$ is equal to $\|\xvec_{u\mid f}(j)\|^2$, as
in \Cref{def:cond}.%

More formally, if we use $\randx_{u}(j)$ to denote the random
indicator variable of $u\in V$ receiving label $j\in \labs$, then
\[
\expcts{}{ \randx_u(j)\mid \text{$f$ is chosen} } = \|\xvec_{u\mid f}(j)\|^2.
\] 
When $\labs = \{0,1\}$, we will drop $j$ and simply use
$\randx_u=\randx_u(1)$.

Note that any such rounding algorithm will be {\bf faithful}
\cite{cdk12}. In particular, for any lifted constraint of the form
$\langle a, \xvec\rangle \le b$,
\[
\expcts{}{ \sum_{u,j} a_{u,j} \randx_u(j)} = \sum_{u,j} a_{u,j}
\|\xvec_{u\mid f}(j)\|^2 \ge b.
\] Combining this with~\Cref{clm:cond-vecs}, we can see that such the
output of such rounding procedure will satisfy each linear constraint
{\bf in expectation}. Furthermore, when labelings are chosen
independently of each other, we can use usual concentration bounds,
such as Chernoff's inequality, to conclude that w.h.p. linear
constraints will be satisfied within some negligible error.

\section{Case Study: Minimum Bisection}
\label{sec:case-study}
All our algorithmic results follow a unified method (except small set
expansion on irregular graphs and unique games, both of which we treat
separately). In this section, we will illustrate the main ideas
involved in our work in a simplified setting, by working out
progressively better approximation ratios for the following basic,
well-studied problem: Given as input a graph $G=(V,E)$ and an integer
size parameter $\mu$, find a subset $U \subset V$ with $|U|=\mu$ that
minimizes the number of edges between $U$ and $V \setminus U$, denoted
$\Gamma_G(U)$. The special case when $\mu = |V|/2$ and we want to
partition the vertex set into two equal parts is the minimum bisection
problem. We will loosely refer to the general $\mu$ case also as
minimum bisection.\footnote{We will be interested in finding a set of
  size $\mu \pm o(\mu)$, so we avoid the terminology {\em Balanced
    Separator} which typically refers to the variant where $\Omega(n)$
  slack is allowed in the set size. In such case, we can do better
  using threshold based rounding (see~\Cref{thm:qcut-approx-guarantee}
  in \Cref{sec:sc}).}
 
For simplicity we will assume
$G$ is unweighted and $d$-regular, however
all our results given in~\Cref{sec:qip} are 
for any weighted undirected graph $G$.
We can formulate this problem as a binary integer programming problem
as follows:
\begin{align}
\label{eq:bisec-qip}
\min_{\xint}\ & 
\sum_{e=\{u,v\}\in E} (\xint_u - \xint_v)^2, \\
\st\  & \sum_u \xint_u = \mu; \quad \xint\in\{0,1\}^{V}\nonumber .
 \end{align} 
 If we let $L$ be the Laplacian matrix for $G$, we can rewrite the
 objective as $\xint^T L \xint$.  We will denote by $\Lnorm$ the normalized
 Laplacian of $G$, $\frac{1}{d} L$.

 Note that the above is a quadratic integer programming (QIP) problem
 with linear constraints. The somewhat peculiar formulation is in
 anticipation of the \lh~semidefinite programming relaxation for
 this problem, which we describe below.

\subsection{SDP Relaxation for Minimum Bisection}
We consider the following relaxation of \cref{eq:bisec-qip}:
\begin{align}
  \label{eq:bisec-sdp}
  \min \quad& \sum_{e=\{u,v\}\in E} \|\xvec_u - \xvec_v\|^2 \\
  \st \quad& \sum_u \xvec_u = \mu \xvec_{\es}, \notag \\
  & \|\xvec_{\es}\|^2 = 1, \notag \\
  & \xmat = [\xvec_S(f)] \in \lasserrei{r}{V}{2} . \notag
\end{align} 
It is easy to see that this is indeed a relaxation of our original QIP
formulation (\ref{eq:bisec-qip}). 

\subsection{Main Theorem on Rounding}
\label{sec:example-rounding}
Let $\xvec$ be an (optimal) solution to the above $r'$-round \lh~SDP.
We will always use $\eta$ in this section to refer to the objective
value of $x$, i.e., $\eta = \sum_{e=\{u,v\}\in E(G)} \left\|x_u -
  x_v \right\|^2$.

Our ultimate goal in this section is to give an algorithm to round the
SDP solution $\xmat$ to a good $\xint \in \{0,1\}^V$ of size very
close to $\mu$, and prove the below theorem.
\begin{theorem}
\label{thm:example-bisec}
For all $r \ge 1$ and $\eps > 0$, by solving $O(r/\eps^2)$-rounds of
\lh~relaxation, we can find $\randx \in \{0,1\}^V$ with
the following properties:
\begin{enumerate}
\item $\randx^T L \randx \le \frac{1+\eps}{\min(1,\lambda_{r+1}(\Lnorm))} \eta$.
\item $\big|\|\randx\|_1 - \mu\big| \le o(1) = O\left(\sqrt{\mu \log(1/\eps)}\right).$
\end{enumerate}
\end{theorem}
Since one can solve the~\lh~relaxation in $n^{O(r')}$ time, we
get the result claimed in the introduction: an $n^{O(r/\eps^2)}$ time
factor $(1+\eps)/\min\{\lambda_r,1\}$ approximation algorithm; the
formal theorem, for general (non-regular, weighted) graphs directly
follows from~\Cref{thm:qip} in \Cref{sec:qip}. Note that if $t =
\argmin_r \{ r \mid \lambda_r(\Lnorm) \ge 1-\eps/2\}$, then this gives
an $n^{O_\eps(t)}$ time algorithm for approximating minimum bisection
to within a $(1+\eps)$ factor, provided we allow $O(\sqrt{n})$
imbalance.

\subsection{The Rounding Algorithm}
\label{sec:rounding-alg-min-bisection}

Recall that the solution $\xmat = [\xvec_S(f)]$ contains a vector
$x_T(f)$ for each $T\in \binom{V}{\le r'}$ and every possible labeling
of $T$, $f\in \{0,1\}^T$ of $T$.  Our approach to round $\xmat$ to
$\xint\in \{0,1\}$ which is an approximate solution of the integer
program \cref{eq:bisec-qip} is similar to the label propagation
approach used in~\cite{akkstv08}.
 
Consider fixing a set of $r'$ nodes, $S\in \binom{V}{r'}$, and
assigning a label $f(s)$ to every $s\in S$ by choosing $f\in
\{0,1\}^S$ with probability $\|\xvec_S(f)\|^2$. (The best choice of
$S$ can be found by brute-forcing over all of $\binom{V}{r'}$, since
solving the SDP takes $n^{O(r')}$ time anyway. But there is
also a faster method to find a good $S$, as mentioned
in~\Cref{thm:column-selection}.) Conditional on choosing a specific
labeling $f$ to $S$, we propagate the labeling to other nodes as
follows: Independently for each $u \in V $, assign $\randx_u \gets 1$
with probability $\|\xvec_{u\mid f}\|^2$ and $\randx_u \gets 0$
otherwise. Observe that if $u\in S$, then $\randx_u = f(u)$
always. Finally, output $\randx$.
Below $\Pi_S$ denotes the projection matrix from~\Cref{not:pis}.
\begin{lemma}
\label{lem:bisec-calc}
For the above rounding procedure, the size of the cut produced,
$\randx^T L \randx$, satisfies:
\begin{equation}
\label{eq:jop-temp}
\expcts{}{\randx^T L \randx} = \eta + \sum_{(u,v)\in E}
\langle \Pi_S^{\perp} \xvec_u, \Pi_S^{\perp} \xvec_v\rangle 
= \eta + \tr\big( \xmat^T \Pi_S^\perp \xmat A \big)
\ .
\end{equation}
Here $A$ is the adjacency matrix of $G$.
\end{lemma}
\begin{proof}
  Note that for $u \neq v$:
  \begin{align*}
    \expcts{}{\randx_u} = & \expcts{f}{\|\xvec_{u\mid f}\|^2} =
    \|\xvec_u\|^2, \\
    \expcts{}{\randx_u \randx_v} = & \sum_f \|\xvec_S(f)\|^2
    \|\xvec_{u\mid f}\|^2 \|\xvec_{v\mid f}\|^2 = \sum_f
    \|\xvec_S(f)\|^2 \frac{\langle \unit{\xvec_S(f)},
      \xvec_u(i)\rangle}{\|\xvec_S(f)\|} \frac{\langle
      \unit{\xvec_S(f)}, \xvec_v(j)\rangle}{\|\xvec_S(f)\|}
    \\
    = &\sum_f {\langle \unit{\xvec_S(f)}, \xvec_u\rangle} {\langle
      \unit{\xvec_S(f)}, \xvec_v\rangle}.
  \end{align*} 
  Since $\{\unit{\xvec_S(f)}\}_f$ is an orthonormal basis, the above
  expression can be written as the inner product of \emph{projections}
  of $\xvec_u$ and $\xvec_v$ onto the span of
  $\{\xvec_S(f)\}_{f:S\to\{0,1\}}$, which we denote by $\Pi_S$. Let us
  now calculate the expected number of edges cut by this rounding.  It
  is slightly more convenient to treat edges $e = \{u,v\}$ as two
  directed edges $(u,v)$ and $(v,u)$, and count directed edges $(u,v)$
  with $\randx_u=1$ and $\randx_v=0$. Therefore,
  \begin{align}
    \expcts{}{\randx^T L \randx}
    =& \sum_{(u,v)\in E} \expcts{}{\randx_u(1-\randx_v)} 
    = \sum_{(u,v)\in E} \Big(\|\xvec_u\|^2 - \langle \Pi_S \xvec_u,
    \Pi_S \xvec_v\rangle\Big) \label{eq:jop0}
    \intertext{$\Pi_S$ is a projection matrix,so $\langle \Pi_S
      \xvec_u , \Pi_S \xvec_v\rangle = \langle \xvec_u ,\xvec_v\rangle -
      \langle \Pi_S^\perp \xvec_u , \Pi_S^\perp
      \xvec_v\rangle$. Substituting this in~\cref{eq:jop0}:}
    =& \sum_{(u,v)\in E}
    \|\xvec_u\|^2
    - \langle \xvec_u, \xvec_v\rangle
    + \langle \Pi_S^{\perp} \xvec_u, \Pi_S^{\perp}
    \xvec_v\rangle\notag\\
    =& ~\eta + \sum_{(u,v)\in E}
    \langle \Pi_S^{\perp} \xvec_u, \Pi_S^{\perp} \xvec_v\rangle.
    \notag
  \end{align}
\end{proof}
Note that the matrix $\Pi_S$ depends on vectors $\xvec_S(f)$ which are
hard to control because we do not have any constraint relating
$\xvec_S(f)$ to a known matrix.  The main driving force behind all our
results is the following fact, which follows since given any $u\in S$
and $i\in \labs$, $\xvec_u(i) = \sum_{f: f(u)=i} \xvec_S(f)$ by SDP
constraints.
\begin{observation} For all $S \in \binom{V}{r'}$,
\[ \mathrm{span}\left(\{\xvec_S(f)\}_{f\in \{0,1\}^S}\right) 
\supseteq \mathrm{span}\left(\{\xvec_u\}_{u\in S}\right) \ .\]
Equivalently for $P_S$ being the projection
matrix onto span of $\{\xvec_u\}_{u\in S}$,
$P_S \preceq \Pi_S$.
\end{observation}
Thus we will try to upper bound the term in~\cref{eq:jop-temp}
by replacing $\Pi_S^\perp$ with $P_S^\perp$. But we cannot directly
perform this switch, as the adjacency matrix $A$ is not PSD. 
\subsection{Factor $1+\frac{1}{\lambda_r}$ Approximation of Cut Value}
\label{sec:simple-anal-bisec}
Our first bound is by directly upper bounding \cref{eq:jop-temp} in
terms of $\|\Pi_S^{\perp} \xvec_u\|^2 \le \|P_S^{\perp}
\xvec_u\|^2$. Using Cauchy-Schwarz and Arithmetic-Geometric Mean
inequalities, \eqref{eq:jop-temp} implies that the expected number of
edges cut is upper bounded by
\begin{equation}
\eta + \frac{1}{2} \sum_{e=(u,v)\in E}
\big(\|\Pi_S^{\perp} \xvec_u\|^2 + \|\Pi_S^{\perp} \xvec_v\|^2\big)
=  \eta + d \sum_u \|\Pi_S^{\perp} \xvec_u\|^2
\le \eta + d \sum_u \|P_S^{\perp} \xvec_u\|^2\ .
\label{eq:jop2}
\end{equation}
If we define the matrix $\ymat \triangleq [\xvec_u]_{u\in V}$, then:
\[
d \sum_u \|P_S^{\perp} \xvec_u\|^2
= d \tr( \ymat^T \ymat_S^\perp \ymat )
= d \|\ymat_S^\perp \ymat\|^2_F.
\]
To get the best upper bound, we want to pick $S \in \binom{V}{r'}$
that minimizes $\|\ymat_S^\perp \ymat\|^2_F$.  It is a well known fact
that among {\em all} projection matrices $M$ of rank $r'$ (not
necessarily restricted to projection onto $\xvec_u$'s), the minimum
value of $\|M^\perp \ymat\|^2_F$ is achieved by matrix $M$ projecting
onto the space of the largest $r'$ singular vectors of
$\ymat$. Further, this minimum value equals $\sum_{i \ge r'+1}
\sigma_i$ where $\sigma_i=\sigma_i(\ymat^T \ymat)$ is $i^{th}$ largest
eigenvalue of $\ymat^T \ymat$. Hence $\|\ymat_S^\perp \ymat\|^2_F \ge
\sum_{i \ge r'+1} \sigma_i$ for every choice of $S$. The following
theorem from~\cite{gs11-svd} shows the existence of $S$ which comes
close to this lower bound:
\begin{theorem}{\cite{gs11-svd}}
\label{thm:column-selection2}
For every real matrix $X$ with column set $V$, and positive integers
$r \le r'$, we have
\[ \delta_{r'}(X) \triangleq \min_{S\in\binom{V}{r'}} \tr(X^T
X_S^{\perp} X) \le \frac{r'+1}{r'-r+1} \Bigl( \sum_{i \ge r + 1}
\sigma_i \Bigr) \ . \]
In particular, for all $\eps \in (0,1)$, $\delta_{r/\eps+r-1} \le
(1+\eps) \Bigl( \sum_{i \ge r+1} \sigma_i \Bigr)$.  Further one can
find a set $S \in \binom{V}{r'}$ achieving the claimed bounds in
deterministic
$O(r n^4)$ 
time.
\end{theorem}
\begin{remark}
  Prior to our paper~\cite{gs11-svd}, it was shown in \cite{bdm11}
  that $\delta_{2 r /\eps + r} \le (1+\eps) \Bigl( \sum_{i \ge r+1}
  \sigma_i \Bigr)$. The improvement in the bound on $r'$ from $2
  r/\eps+r$ to $r/\eps+r-1$ to achieve $(1+\eps)$ approximation is not
  of major significance to our application, but since the tight bound
  is now available, we decided to state and use it.  {\hfill $\Box$}
\end{remark}
Picking the subset $\bestSi{} \in \binom{V}{r'}$ that achieves the
bound guaranteed by \Cref{thm:column-selection2}, we have
\[ \|\ymat_S^\perp \ymat\|^2_F \le (1+\eps) \sum_{i > r} \sigma_i \
. \]
In order to relate this quantity to the SDP objective value $\eta =
\tr(\ymat^T \ymat L)$, we use the fact that $\tr(\ymat^T \ymat L)$ is
minimized when eigenvectors of $\ymat^T \ymat$ and $L$ are matched in
reverse order: $i^{th}$ largest eigenvector of $\ymat^T \ymat$
coincides with the $i^{th}$ smallest eigenvector of $L$.  Letting
$0=\lambda_1(\Lnorm)\le \lambda_2(\Lnorm) \le \ldots \le
\lambda_n(\Lnorm) \le 2$ be the eigenvalues of normalized graph
Laplacian matrix, $\Lnorm = \frac{1}{d} L$, we have
\[
\frac{\eta}{d} = \frac{1}{d} \tr(\ymat^T \ymat L) \ge \sum_{i} \sigma_i \lambda_i(\Lnorm)
\ge \sum_{i\ge r+1} \sigma_i \lambda_{r+1}\ge 
(1+\eps)^{-1} 
\lambda_{r+1}(\Lnorm) \|\ymat_S^\perp \ymat\|^2_F.
\]
Plugging this into~\cref{eq:jop2}, we can conclude our first bound:
 \begin{theorem}
 \label{thm:minbisec-first}
 The rounding algorithm given in
 \Cref{sec:rounding-alg-min-bisection} cuts at most
 \[ \left(1+\frac{1+\eps}{\lambda_{r+1}(\Lnorm)} \right) 
 \sum_{e=(u,v)\in E} \|\xvec_u - \xvec_v\|^2 \]
 edges in expectation.
 In particular, the algorithm cuts at most a factor
 $1+\frac{1+\eps}{\lambda_{r+1}(\Lnorm)}$ more edges than the optimal
 cut with $\mu$ nodes on one side.\footnote{We will later argue that
   the cut will also meet the balance requirement up to $o(\mu)$
   vertices.}
\end{theorem}
Note that $\lambda_n(\Lnorm)\le 2$, hence even if we use $n$-rounds
(in which case $\xmat$ is an integral solution), the smallest upper
bound we can show is $\frac{3}{2}$.  Although this is too weak by
itself for our purposes, this bound will be crucial to obtain our
final bound.
\subsection{Improved Analysis and Factor $\frac{1}{\lambda_r}$
  Approximation on Cut Value}
\label{sec:simple-anl-bisec}
First notice that \cref{eq:jop-temp} can be written as:
\begin{equation}
\label{eq:jop4}
\expcts{}{\mbox{number of edges cut}} = 
\tr( \xmat^T \xmat L) + 
\tr( \xmat^T \Pi_S^\perp \xmat A)
= d \tr( \xmat^T \Pi_S^\perp \xmat ) + \tr( \xmat^T \Pi_S \xmat L )\ .
\end{equation} 
If value of this expression is larger than 
$(1+\eps) \frac{\eta}{ \lambda_{r+1}} + \eta \eps$, 
then value of $\tr( \xmat^T \Pi_S^{} \xmat L)$
has to be larger than $\eps \eta$ due to 
the bound we proved on $\tr( \xmat^T \Pi_S^{\perp} \xmat)$.
Consider choosing another subset $T$ that achieves 
the bound $\delta_r(\Pi_S^{\perp} \xmat)$. The crucial 
observation is that distances between neighboring
nodes on vectors $\Pi_S^{\perp} X$
has decreased by an additive factor of $\eta \eps$,
\[
\tr(\xmat^T \Pi_S^{\perp} \xmat L) = 
\tr(\xmat^T \xmat L) - \tr(\xmat^T \Pi_S^{} \xmat L) < \eta(1 - \eps)
\] so that
$ \tr(\xmat^T \Pi_{S\cup T}^{\perp} \xmat)< (1-\eps) \frac{(1+\eps)
  \eta}{\lambda_{r+1}}$.  
Now, if we run the rounding algorithm with $S \cup T$ as the seed set,
and (\ref{eq:jop4}) with $S \cup T$ in place of $S$ is larger than
$\frac{(1+\eps) \eta}{\lambda_{r+1}} + \eta \eps$, then $\tr(\xmat^T
\Pi_{S\cup T} \xmat L) > 2 \eps \eta$. Hence
\[ \tr(\xmat^T \Pi_{S \cup T}^{\perp} \xmat L) \le \tr(\xmat^T \xmat L) - \tr(\xmat^T
\Pi_{S\cup T} \xmat L) < \eta (1-2\eps) \ . \] 
Picking another set $U$, we will have $\tr(\xmat^T \Pi_{S\cup T\cup
  U}^{\perp} \xmat)< (1- 2 \eps) \frac{(1+\eps)
  \eta}{\lambda_{r+1}}$. Continuing this process, if the quantity
\cref{eq:jop4} is not upper bounded by $\frac{(1+\eps) \eta}{
  \lambda_{r+1}} + \eta \eps$ after $\lceil \frac{1}{\eps}\rceil$ many
such iterations, then the total projection distance becomes:
\[ 
0 \le
\tr(\xmat^T \Pi_{S\cup T\cup \ldots}^{\perp} \xmat)<
(1- \lceil1/\eps\rceil \eps)
\frac{(1+\eps)\eta}{\lambda_{r+1}} \le 0
\] which is a contradiction. For formal statement and proof in a more general setting,
see~\Cref{thm:bounding-delta-r} in \Cref{sec:seed-good}.
\begin{theorem}
\label{thm:minbisec-ptas-style}
The expected number of edges cut by the rounding algorithm
from~\Cref{sec:rounding-alg-min-bisection} using the seed selection
procedure as described in \Cref{sec:simple-anl-bisec} is at most
$(1+\eps)/\min\{1,\lambda_{r+1}(\Lnorm)\}$ times the size of the
optimal cut with $\mu$ nodes on one side. Here $\lambda_{r+1}(\Lnorm)$
is the $(r+1)$'th smallest eigenvalue of the normalized Laplacian
$\Lnorm = \frac{1}{d} L$ of the $G$.
\end{theorem}
\subsection{Bounding Set Size}
We now analyze the balance of the cut, and show how to ensure that
$\|\randx\|_1 = \mu \pm o(\mu)$ in addition to $\randx^T L \randx$
being close to the expected bound of \Cref{thm:minbisec-ptas-style}
(and similarly for \Cref{thm:minbisec-first}).

Let $f: \bestSi \to \{0,1\}$ be fixed. We will show that conditioned
on finding cuts with small cost, the probability that one of them has
size $\approx \mu$ is bounded away from zero. We can then use a simple
Markov bound to show that there is a non-zero probability that both
cut size and set size are within $O(1)$-factor of corresponding
bounds. But by exploiting the independence in our rounding algorithm
and \lh~relaxations of linear constraints, we can do much better.
Note that in the $r'$-round \lh~relaxation, we had:
\[
\sum_u \xvec_u = \mu \xvec_{\es}.
\]
This means, for each $f\in \{0,1\}^{\bestSi{}}$:
\[
\sum_u \|\xvec_{u\mid f}\|^2 = \mu\ .
\] 
This implies that conditioned on the choice of $f$, the expectation of
$\sum_u \randx_u$ is $\mu$ and $\xint_u$ for various $u$ are all
independent. Applying the Chernoff bound, we get
\[ \probs{} {~\Bigl|\sum_u \randx_u - \mu\Bigr| \ge 2 \sqrt{\mu
    \log\frac{1}{\zeta}}} \le o(\zeta) \le \frac{\zeta}{3}.\]
Consider choosing $f\in\{0,1\}^{\bestSi{}}$ so that ${\mathbb
  E}\bigl[\mbox{number of edges cut} ~ \vline ~ f\bigr] \le {\mathbb
  E}\bigl[\mbox{number of edges cut}\bigr] \triangleq b$.
By Markov inequality, if we pick such an $f$,
$\mathsf{Pr}\bigl[\mbox{number of edges cut}\ge (1+\zeta)b\bigr] \le 1
- \frac{\zeta}{2}$, where the probability is over the random
propagation once $\bestSi{}$ and $f$ are fixed.
Hence with probability at least $\frac{\zeta}{6}$, the subset found
has cut cost $\le (1+\zeta) b$ and size in the range $\mu \pm 2
\sqrt{\mu \log\frac{1}{\zeta}}$. Taking $\zeta = \eps$ and repeating
this procedure $O\left(\eps^{-1} \log n\right)$ times, we get a high
probability statement and %
finish~\Cref{thm:example-bisec} on minimum bisection.

\section{Analysis of Propagation Rounding}
\label{sec:prop-round}
Intuitively, the main quantity that affects our approximation factor
is how far $\|\xvec_{u\mid f}\|^2$ is from $\{0,1\}$ in expectation
over $f$. In this section, we will show how to choose seed sets so as
to relate this distance to certain spectral properties of the
underlying graph. We start with proving some simple properties:

Let $\randx_{A\mid f}(g)$ be a random variable over $\{0,1\}$ with
$\probs{}{\randx_{A\mid f}(g)=1}= \|\xvec_{A\mid f}(g)\|^2$. Given
fixed $f_0\in \labs^{S_0}$ and subset $S$, we choose our next
conditioning $f:S\to \labs$ by sampling $f$ with probability
$\|\xvec_{S\mid f_0}(f)\|^2$. We will denote this by $f \sim
\|\xvec_{S\mid f_0}(f)\|^2$.

\begin{claim}
  \label{thm:expct-over-f-single}
  \[
  \expcts{f \sim \|\xvec_{S}(f)\|^2}{ \expcts{}{
      \randx_u(j)\mid \text{$f$ is chosen}} } = \|\xvec_{u}(j)\|^2.
  \]
\end{claim}
\begin{proof}
  \[
  \expcts{f \sim \|\xvec_{S}(f)\|^2}{ \expcts{}{
      \randx_u(j)\mid \text{$f$ is chosen}} } = 
  \sum_f \|\xvec_{S}(f)\|^2 \frac{\langle \xvec_S(f),
    \xvec_u(j)\rangle}{\|\xvec_S(f)\|^2}
  = \langle \xvec_{\es}, \xvec_u(j)\rangle = \|\xvec_u(j)\|^2.
  \tag*{\qedhere}
  \]
\end{proof}

After having chosen $S$ and $f$, there are two natural rounding
schemes:
\begin{enumerate}
\itemsep=1ex
\item {Independent Rounding:} For each $u\in V$, choose $j\in \labs$
  with probability $\|\xvec_{u\mid f}(j)\|^2$. This rounding scheme
  will be used for unique games type problems as well as independent
  set. 
\item {Threshold Rounding:} (When $k=2$) Choose $\tau \in (0,1]$ uniformly at
  random. For each $u\in V$, let $\randx_u \gets 1$ if $\|\xvec_{u\mid
    f}(j)\|^2 \ge \tau$ and $\randx_u \gets 0$ otherwise. We will use
  this rounding for minimum cut type problems such as sparsest cut,
  balanced separator, etc.
\end{enumerate}
\subsection{Independent Rounding Scheme}
\label{sec:independent-rounding}
As we discussed in~\Cref{sec:rounding-overview}, it is easy to show
that, under such scheme, linear constraints will be satisfied with
high probability within error $\widetilde{O}(\sqrt{n})$ (here
$\widetilde{O}$ hides logarithmic terms) using concentration
inequalities. Therefore we only need to bound the probability that
some constraint on a pair of variables $(u,v)$ is left unsatisfied.
\begin{claim}
  \label{thm:expct-over-f-binary}
  Given $u\neq v\in V$ and $i,j\in\labs$,
  \[
  \expcts{f}{\randx_u(i) \randx_v(j)} = \xvec_u(i)^T \Pi_S \xvec_v(j).
  \]
\end{claim}
\begin{proof}
  \begin{align*}
    \expcts{f}{\randx_u(i) \randx_v(j)} = & 
    \sum_f \|\xvec_S(f)\|^2 \langle \dunit{\xvec_S(f)},
    \xvec_u(i)\rangle
    \langle \dunit{\xvec_S(f)}, \xvec_v(j)\rangle
    = \sum_f \langle \unit{\xvec_S(f)},
    \xvec_u(i)\rangle
    \langle \unit{\xvec_S(f)}, \xvec_v(j)\rangle \\
    = & \xvec_u(i)^T \Big( \sum_f \unit{\xvec_S(f)}\cdot
    \unit{\xvec_S(f)}^T 
    \Big) \xvec_v(j) =  \xvec_u(i)^T \Pi_S \xvec_v(j).
  \end{align*} In the last identity, we used~\Cref{not:pis}.
\end{proof}
\begin{lemma}
  \label{thm:ind-expct-over-mtx}  
  Given set of seeds $S$, for any symmetric matrix $A$ of appropriate
  dimensions:
  \[
  \expcts{}{\sum_{u,v} \sum_{i,j} \randx_u(i) \randx_v(j)
    A_{(u,i),(v,j)}} = \tr(\xmat^T \Pi_S \xmat A) + \tr(\xmat^T
  \Pi_S^{\perp} \xmat \diag(A)).
  \]
\end{lemma}
\begin{proof}
  Fix $u$ and $i$. Since $\Pi_S$ is a projection matrix, we have:
  $
  \expcts{}{\randx_u(i) } 
  =
  \|\xvec_u(i)\|^2 = \xvec_u(i)^T \Pi_S \xvec_u(i) + \xvec_u(i)^T
  \Pi_S^\perp \xvec_u(i)$.   
  For $A^o \triangleq A - \diag(A)$ (the
  off-diagonal entries of $A$):
  \begin{align*}    
    \expcts{}{\randx^T A \randx}
    = & \expcts{}{\randx^T \diag(A) \randx}
    + \expcts{}{\randx^T A^o \randx} \\
    = & \tr(\xmat^T \Pi_S^{\perp} \xmat \diag(A)) 
    + \tr(\xmat^T \Pi_S^{} \xmat \diag(A))
    + \tr(\xmat^T \Pi_S^{} \xmat A^o)\\
    = & \tr(\xmat^T \Pi_S^{\perp} \xmat \diag(A)) 
    + \tr(\xmat^T \Pi_S^{} \xmat A)
  \end{align*}
  where we used~\Cref{thm:expct-over-f-binary} in the first step.
\end{proof}
Observe that $\Pi$ is a matrix that depends on the higher-order
vectors $\{\xvec_S(f)\}$. This makes it very difficult to reason about
$\Pi$ matrix. In~\Cref{thm:pi-to-p}, we will relate the matrix $\Pi_S$
to projection matrices onto the linear subspaces of
$\{\xvec_{S}(f)\}_f$'s span:
\begin{lemma}\label{thm:pi-to-p}
  For any $S\subseteq [n]$, let $Y$ be a
  matrix whose columns are linear combinations of vectors from the set
  $\left\{ \xvec_{S}(g) \right\}_{g}$. Then:
  \begin{enumerate}
  \item $\Pi_{S}$ is a projection matrix, i.e. $\Pi_{S}^2= \Pi_{S}$;
  \item $\Pi_{S}^\perp$ is the projection matrix onto the
    orthogonal complement of $\Pi_S's$ row span, i.e. $\Pi_S^\perp \Pi_S =
    0$;
  \item $\Pi_S$ dominates $Y^{\Pi}$, i.e. $\Pi_{S} \succeq
    Y^{\Pi}$ where $Y^{\Pi}$ is the projection matrix onto the span of
    $Y$.
  \end{enumerate}
\end{lemma}
\begin{proof}
Denote $\Pi = \Pi_S$ for notational simplicity. 
  By construction, $\Pi$ is symmetric and $\Pi\succeq 0$.  We will
  show that $\Pi^2= \Pi$ which implies $\Pi$ is a projection
  matrix:
  \[
  \Pi^2 = \sum_{g, h} \frac{1}{\|\xvec(g)\|^2 \xvec(h)\|^2} \xvec(g)
  \xvec(g)^T \xvec(h) \xvec(h) = \sum_g \unit{\xvec(g)}
  \unit{\xvec(g)}^T = \Pi.\] Since $Y$'s columns are linear
  combinations of $\Pi$, this means $\Pi Y = \Pi$. Using the fact that
  $0 \preceq Y^{\Pi}\preceq I$, we obtain $Y^{\Pi} = \Pi Y^{\Pi} \Pi
  \preceq \Pi \cdot I \cdot \Pi = \Pi$.
\end{proof}
%
%
\subsection{Threshold Based Rounding Scheme}
\label{sec:threshold-rounding}
We will use such rounding for problems that involve {\em minimizing}
the number of edges cut under various constraints involving quantities
like partition size or the number of edges cut in some other
``demand'' graph. Our first quantity of interest is the probability of
separating pair $(u,v)$.
\begin{claim}
  \label{thm:threshold-ub-0}
  $\big| \| \xvec_{u\mid f}\|^2 - \| \xvec_{v\mid f}\big\|^2 \big| 
  \le \|\xvec_{u\mid f} - \xvec_{v\mid f}\|^2.
  $
\end{claim}
\begin{proof}
  \begin{align*}
    \big| \| \xvec_{u\mid f}\|^2 - \| \xvec_{v\mid f}\big\|^2 \big| 
    & = \big| \| \xvec_{uv\mid f}(10)\|^2 - \| \xvec_{uv\mid
      f}(01)\big\|^2 \big| \\
    & \le \| \xvec_{uv\mid f}(10)\|^2 + \| \xvec_{uv\mid
      f}(01)\big\|^2  = \|\xvec_{u\mid f} - \xvec_{v\mid f}\|^2.  \tag*{\qedhere}
  \end{align*}
\end{proof}
\begin{lemma}
  \label{thm:threshold-ub}
  For fixed $f$, let $\tau$ be chosen uniformly at random from
  $(0,1]$. For all $u\in V$, if we let $\randx_u$ be the indicator
  of $\|\xvec_{u\mid f}\|^2 \ge \tau$, then the probability
  of $\randx_u\neq \randx_v$ is bounded as follows:
  \[
  \probs{\tau}{\randx_u \neq \randx_v \mid\ f} \le {\|\xvec_{u\mid f} -
  \xvec_{v\mid f}\|^2}.
  \]
  If $f$ is chosen with probability $\|\xvec(f)\|^2$, then:
  \[
  \probs{f,\tau}{\randx_u \neq \randx_v} \le {\|\xvec_{u} -
  \xvec_{v}\|^2}.
  \]
\end{lemma}
\begin{proof}
  First part follows %
  from~\Cref{thm:threshold-ub-0}. Second part follows
  from~\Cref{clm:cond-vecs}(\ref{thm:cond:2}).
\end{proof}
For (non-uniform) sparsest cut problem, we need the following
lower bound:
\begin{claim}
  \label{thm:threshold-lb}
  Suppose the threshold, $\tau$, is chosen uniformly at random from
  $(0,1]$. Then, the probability of $\randx_u \neq \randx_v$ over
  random $f$ is:
    \[
    \probs{f,\tau}{\randx_u \neq \randx_v\ \mid\ f} \ge \|\xvec_u -
    \xvec_v\|^2- \|\Pi_S^{\perp} (\xvec_u - \xvec_v)\|^2.
  \]
\end{claim}
\begin{proof}
  \begin{align*}
    \probs{}{\randx_u \neq \randx_v\ \mid\ f} 
    = & \expcts{f}{ 
      \big| \langle \dunit{\xvec_S(f)} , \xvec_u \rangle- \langle
      \dunit{\xvec_S(f)} , \xvec_u \rangle \big|} 
    \ge \expcts{f}{ 
      \big( \langle \dunit{\xvec_S(f)} , \xvec_u \rangle- \langle
      \dunit{\xvec_S(f)} , \xvec_u \rangle \big)^2} \\
    = & \left\| \Pi_S \left( \xvec_u - \xvec_v\right) \right\|^2.
  \end{align*}
  Using the fact that $\Pi_S$ is a projection
  matrix~(see~\Cref{thm:pi-to-p}) with orthogonal complement
  $\Pi_S^\perp$ finishes the proof.
\end{proof}

\section{Choosing A Good Seed Set by Column Selection}
\label{sec:seed-good}
As we have seen in \Cref{sec:case-study,sec:prop-round}, the
performance depends on minimizing $\tr(\xmat^T \Pi_{S}^\perp \xmat
B)$. In this section, we will relate this quantity to the spectrum of
objective matrix via column based matrix reconstruction.
First, we need a generalized form of von Neumann's Trace
Inequality~\cite{hj-mat-book} for relating the approximation factor to
the (generalized) spectrum.
\begin{theorem}[Generalized Trace Inequality]
  \label{thm:gen-trace-ineq}
  Let $A, B \in \sympm^{C}$ be two matrices with 
  $B = \Gamma^T \Gamma$ for some $\Gamma \in \R^{R,C}$ such that
  $\nll(B) \subseteq \nll(A)$. Then for any $X \in \sympm^C$, the following bound holds:
  \begin{equation}
    \label{eq:gen-trac-ineq-1}
    \sum_{j} \sigma_j(\Gamma X \Gamma^T) \lambda_j(A , B) \le 
    \tr( X A),
  \end{equation}
  where $\sigma_j$,$\lambda_j$ are as defined
  in~\cref{eq:sigma-i-def,eq:glambda-i-def}, respectively.
  Moreover for any positive integer $r$,
  \begin{equation}
    \label{eq:gen-trac-ineq-2}
    \sum_{j\ge r} \sigma_j(\Gamma X \Gamma^T) \le 
    \frac{\tr( X A)  }{\lambda_{r}(A, B)}.    
  \end{equation}
\end{theorem}
\begin{proof} Unless noted otherwise, we will use $\sigma_j=\sigma_j(\Gamma
  X \Gamma^T)$ and $\lambda_j=\lambda_j(A,B)$.
  Given \cref{eq:gen-trac-ineq-1}, we can derive
  \cref{eq:gen-trac-ineq-2} easily as follows. Since $0\le \lambda_1 \le
  \lambda_2 \le \ldots \le \lambda_r$:
  \[
  \tr(X A) \ge 
  \sum_{j} \sigma_j \lambda_j \ge \sum_{j\ge r} \sigma_j \lambda_j
  \ge \sum_{j\ge r} \sigma_j \lambda_r = \lambda_r \sum_{j\ge r} \sigma_j.
  \]  
  Now we will prove \cref{eq:gen-trac-ineq-1}. Since $\nll(A)
  \supseteq \nll(B)$, $B^{\dagger} B A = A B^{\dagger} B = A$. Hence:
  \begin{align*}
    \tr(X A) = & \tr(X B^{1/2} B^{\dagger/2} A B^{\dagger/2} B^{1/2}) 
    = \tr( B^{1/2} X B^{1/2} B^{\dagger/2} A B^{\dagger/2}) 
    \intertext{For $q_i$'s being the eigenvectors of $B^{\dagger/2} A
      B^{\dagger/2}$, we have $B^{\dagger/2} A
      B^{\dagger/2} = \sum_j \lambda_j q_{j+r} q_{j+r}^T$ where $r =
      \rank(B)$. Substituting this into previous equation, we get:}
    = & \sum_{j=1}^{n-r} \lambda_j q_{j+r}^T B^{1/2} X B^{1/2} q_{j+r}
    = \sum_{j=1}^{n-r} \lambda_j q_{j+r}^T B^{1/2} B^{\dagger} \Gamma^T \Gamma X \Gamma^T
    \Gamma B^{\dagger} B^{1/2} q_{j+r} \\
    = & \sum_{j=1}^{n-r} \lambda_j q_{j+r}^T B^{\dagger/2} \Gamma^T Y %
    \underbrace{\Gamma B^{\dagger/2} q_{j+r}}_{\triangleq p_{j}}
    = \sum_j \lambda_j p_j^T Y p_j
    \ge \sum_j (\lambda_i)\!\uparrow_j (p_i^T Y p_i\mid i)\!\downarrow_j
    \intertext{Assume $p_j$'s were all orthonormal. Then the
      sequence $(p_j^T Y p_j)_j$ is majorized by
      $(\sigma_j(Y))=(\sigma_j(\Gamma X \Gamma^T))_j$. Since $f(x) = \sum_j
      (\lambda_i)\!\uparrow_j x\!\downarrow_j$ is Schur-concave, for any $y$ that
      majorizes $x$, $f(x) \ge f(y)$:}
    \ge& \sum_j (\lambda_i)\!\uparrow_j \sigma_j = \sum_j \lambda_j \sigma_j.
  \end{align*}
  We will finish the proof by proving that $p_i$'s are orthonormal:
  \begin{align*}
    \langle p_i, p_j\rangle = & q_{i'}^T B^{\dagger/2} \Gamma^T \Gamma B^{\dagger/2} q_{j'} =
    q_{i'}^T B^{\dagger/2} B B^{\dagger/2} q_{j'} = q_{i'}^T
    B^{\dagger} B q_{j'} = q_{i'}^T q_{j'}
    =
    \begin{dcases*}
      1 & if $i=j$, \\
      0 & else.
    \end{dcases*} 
\qedhere
  \end{align*}
\end{proof}
The following theorem says that given a matrix the sum of whose 
largest $r$ eigenvalues is large, we can always find a set of $O(r)$
columns that approximates the matrix well (in Frobenius norm). A bound
of the form $1+O(r/r')$ is also given in~\cite{bdm11}.
\begin{theorem}{\cite{gs11-svd}}
\label{thm:column-selection}\label{thm:svd}
For any $X \in \R^{R,C}$ with column set $C$,  and positive
integers $r \le r'$, we have
\[ \delta_{r'}(X) \triangleq \min_{S\in\binom{C}{r'}}
\|X_{S}^{\perp} X \|^2_F \le \frac{r'+1}{r'-r+1} \Bigl(
\sum_{i \ge r + 1} \sigma_i \Bigr) \ . \] In particular, for all $\eps
\in (0,1)$,
$\delta_{r/\eps+r-1} \le (1+\eps) \sum_{i \ge r+1} \sigma_i$.  
\end{theorem}
We combine \Cref{thm:gen-trace-ineq,thm:svd} in the following corollary.
\begin{corollary}
\label{lem:markov-bound}
Given positive integers $r, w$ and $\eps \in (0,1)$
the following holds.
For any 
$\xmat = [\xvec_S(f)] \in \lasserrei{r'}{V}{k}$ where 
\[
r' = w \left(\frac{r}{\eps} + r - 1\right),
\]
and 
$A, B\in \sympm^{C}$ with $\nll(B)\subseteq \nll(A)$, 
provided
that $B=\Gamma^T \Gamma$ for some $\Gamma$ of width-$w$ 
there exists $\bestS \in \binom{V}{r'}$ such that:
\begin{equation}
\tr( \xmat^T \Pi_{\bestS}^{\perp} {\xmat} B) 
\le
(1+\eps)
\frac{\tr(\xmat^T \xmat A)}{\lambda_{r+1}(A,B)}.
\label{eq:delta-r:markov-bound-0}
\end{equation} 
For any $T$, if we replace $\xmat$ and $\bestS$ with $\Pi_T^\perp
\xmat$ and $\bestS \cup T$, respectively, then the claim still holds.
\end{corollary}
\begin{proof}
Let $\ymat \triangleq %
\xmat
\Gamma^T$. By~\cref{eq:gen-trac-ineq-2} of~\Cref{thm:gen-trace-ineq},
we see that
\[
\sum_{j\ge r+1} \sigma_j(\ymat^T \ymat) \le \frac{ \tr(\xmat^T
  \xmat
  A) }{\lambda_{r+1}(A;B)}.
\]
Consequently, for $r'' \triangleq O(r/\eps)$, we can
apply~\Cref{thm:svd} to infer the existence of $r''$ columns, $S$, of
$\ymat$, such that:
\[
\|\ymat_S^\perp \ymat\|^2_F \le (1+\eps)\frac{ \tr(\xmat^T
  \xmat
  A) }{\lambda_{r+1}(A;B)}.
\]
Since each column of $\ymat$ is a linear combination of $w$ vectors
from $\xmat$, there exists a set $S'$ of size $|S'| \le |S| \cdot w
= r'' w$ such that each column of $\ymat_{S'}$ is a linear combination
of $%
\xmat_{S'}$. Finally we let $\bestS$ be 
the set of all variables which appear in some $f
\in S'$. Since $\Pi_{\bestS} \xvec(f) = \xvec(f)$ whenever $f :
T\to\labs$ for some $T \subseteq \bestS$, this implies
$\Pi_{\bestS} \succeq \ymat_{S}^\proj$. Consequently:
\[
\tr(\xmat^T \Pi_{\bestS}^\perp \xmat B)= \| \Pi_{\bestS}^\perp \ymat \|^2_F 
\le \| \ymat_{S}^\perp \ymat\|^2_F \le (1+\eps) \frac{ \tr(\xmat^T
  \xmat A)}{\lambda_{r+1}(A;B)} \ . \qedhere
\]
\end{proof}
The following theorem will be our main result in this section.
\begin{theorem}[Main technical theorem]
\label{thm:bounding-delta-r}
Given positive integers $r, w$ and $\eps \in (0,1)$
the following holds.
For any 
$\xmat = [\xvec_S(f)] \in \lasserrei{r'}{V}{k}$ where 
\[
r' \le \frac{8 r w}{\eps^2},
\]
and $A, B\in \sympm^{C}$ with $\nll(B)\subseteq \nll(A)$, provided that $B=\Gamma^T \Gamma$ for some $\Gamma$ of width-$w$, there exists $\bestS \in \binom{V}{r'}$ with:
 \begin{equation}
\tr( \xmat^T \Pi_{\bestS}^{\perp} {\xmat} B) 
+ 
\tr({\xmat}^T \Pi_{\bestS}^{} {\xmat} A)
<
(1+\eps)
\frac{\tr(\xmat^T \xmat A)}{\min\{\lambda_{r+1}(A,B),1\}}.
\label{eq:delta-r:seed-bound}
\end{equation} 
\end{theorem}
\begin{remark}
  \label{rem:diag-b-low-width}
  We will mainly use~\Cref{thm:bounding-delta-r} with $B$ being a
  diagonal matrix with non-negative entries on the diagonal. In such case, the diagonal matrix $\Gamma$ formed
  by taking the element-wise square root $B$ has width $1$ and
  satisfies $B = \Gamma^T \Gamma$. 
\end{remark}
\begin{proof} 
  Consider picking the ``seed'' nodes $\bestS$ in the following iterative
  way for each $i\in\{1,2,\ldots\}$. At iteration $i$, let $S(i)$ be
  the set of all seeds picked so far (initially $S(0)\gets \es$) and
  $\xmat(i+1) \gets \Pi_{S(i)}^\perp \xmat$ (initially $\xmat(0) \gets
  \xmat$). We use \Cref{lem:markov-bound} on the matrix $\xmat(i)$ to
  choose $4 r w/\eps$ additional columns, $\Delta S(i+1)$.  Finally we
  set $S(i+1) \gets S(i) \cup \Delta S(i+1)$ and keep iterating
  until~\cref{eq:delta-r:seed-bound} is satisfied.  Note that $|S(i)|
  \le O\left( \frac{i w r}{\eps} \right)$. Thus it suffices to show
  that this procedure will stop after $ \frac{1}{\eps}$ iterations.

  Let $\lambda' \triangleq \min(\lambda_{r+1},1)$ and define 
  \[\delta_i \triangleq \tr( \xmat(i)^T {\xmat(i)}
  B),\qquad \eta_i \triangleq \tr( \xmat(i)^T {\xmat(i)} A), \qquad
  \eta \triangleq \tr( \xmat^T {\xmat} A),\qquad \xi_i \triangleq
  \delta_i + \eta - \eta_i.\] One can easily verify that $\eta =
  \eta_0 \ge \eta_{i} \ge \eta_{i+1} \ge 0$, $\delta_{i} \ge
  \delta_{i+1}$ and
  \[
  \tr(\xmat^T \Pi_{S(i)} \xmat A) = \eta - \eta_i
  \implies 
  \tr(\xmat^T \Pi_{S(i)}^\perp \xmat B) + \tr(\xmat^T \Pi_{S(i)} \xmat
  A) = \xi_i.
  \]
  Our goal is to show that, %
  if the procedure did not stop at $i^{th}$
  iteration, then:
  \[
  \eta - \eta_i \ge i \eps \eta / 2.
  \] 
  For $i > \frac{1}{\eps}$, this yield the desired contradiction,
  meaning that the procedure stopped after that many iterations.
  In order to lower bound $\eta - \eta_i$, we first recall the bound
  from~\Cref{lem:markov-bound}:
  \begin{align}   
    \label{eq:thm-bounding-delta-r-0}
    \delta_{i+1} \le &
    (1+{\eps}/{2}) \frac{\eta_i}{\lambda_{r+1}}
    \le 
    (1+{\eps}/{2}) \frac{\eta_i}{\lambda'}. 
    \intertext{ Substituting $\xi_{i+1} \ge (1+\eps)
      \frac{\eta}{\lambda'}$ in~\cref{eq:thm-bounding-delta-r-0}:}       
    \label{eq:thm-bounding-delta-r-1}
    (1+\eps) \frac{\eta}{\lambda'} \le\ & \xi_{i+1} = \delta_{i+1} + \eta -
    \eta_{i+1}  
    \le (1+\eps/2) \frac{\eta_i}{\lambda'} + \eta - \eta_{i+1}. 
    \intertext{After rearranging terms in
      \cref{eq:thm-bounding-delta-r-1},
      our proof is complete:}
    \eta - \eta_{i+1} &\ge \left(\frac{1+\eps/2}{\lambda'}\right) (\eta-\eta_i) +
    \frac{\eps}{2\lambda'} \eta 
    \ge (1+\eps/2) (\eta-\eta_i) +
    (\eps/2) \eta 
    \notag \\
    & \qquad \vdots\notag \\
    & \ge \frac{\eps i}{2} \eta.\tag*{\qedhere}
  \end{align}
\end{proof}

\section{Algorithms Based on Independent Rounding} 
\label{sec:app-independent}
In this section, we will present approximation algorithms which are
based on independent rounding scheme. Unless stated otherwise, our
algorithms will all follow the same outline:
\begin{enumerate}
\item For some matrix $A$, choose seeds $\bestS$ as described
  in~\Cref{thm:bounding-delta-r} with $B \triangleq \diag(A)$.
\item Sample $f:\bestS \to \labs$ with probability $\|\xvec(f)\|^2$. 
\item For each $u\in V$, independently sample a label $j\in \labs$
  from the distribution $(\|\xvec_{u\mid f}(j)\|^2)$.
\end{enumerate}

\subsection{Quadratic Integer Programming}
\label{sec:qip}
\begin{theorem}
  \label{thm:qip}
  Consider a quadratic integer programming problem with objective
  function of the form
  \[
  \xint^T A \xint 
  \] for some $A\in \sympm^{V\times \labs}$ subject to various linear
  constraints. Then by solving $O(r/\eps^2)$-rounds of \lh~relaxation,
  we can obtain a solution $\xint$ such that:
  \begin{itemize}
  \item $\xint^T A \xint \le (1+\eps)
    \frac{\opt}{\lambda_{r}(A,\diag(A))}.$
  \item $\xint$ is obtained by a faithful and independent rounding. 
    In particular, concentration bounds hold for linear constraints.
  \end{itemize}
\end{theorem}
\begin{remark}[Faster Implementation]
  \label{rem:faster-qip}  
  By using the faster solver from~\cite{gs12-fast}, we can implement a
  matching algorithm for~\Cref{thm:qip} whose running time is
  $2^{O(r/\eps^3)} n^{O(1/\eps)}$.
\end{remark}
\begin{proof}
  For fixed $\bestS$, the expected value of the rounding is equal to 
  \[
  \tr(\xmat^T \Pi_{\bestS} \xmat A) + 
  \tr(\xmat^T \Pi_{\bestS}^\perp \xmat \diag(A))
  \] by~\Cref{thm:ind-expct-over-mtx}. If we choose $\bestS$ using
  \Cref{thm:bounding-delta-r} with $B \gets \diag(A)$, then this value
  is upper bounded by:
  \[
  (1+\eps) \frac{\tr(\xmat^T \xmat A)}{\lambda_{r}(A,\diag(A))}. \tag*{\qedhere}
  \]
\end{proof}

\subsection{Maximum Cut and Unique Games}
\label{sec:ug}
\label{sec:ug-proof}
\def\VECX{X}
\def\lassx{X}%
\def\sym{\mathrm{Sym}}
In this section, we obtain our algorithmic result for Unique Games type
problems.
Let us quickly recall the definition of the Unique Games problem. 
An instance of Unique Games consists of a graph $G=(V,C)$, 
a label set $[k]$, and bijection constraints $\pi_{uv} : [k]\to [k]$ for 
each edge $\{u,v\}$. The goal is to find a labeling $f : V \to [k]$ 
that minimizes the number of unsatisfied constraints, 
where $e=\{u,v\}$ is unsatisfied if $\pi_{uv}(f(u))\neq f(v)$ 
(we assume the label of the lexicographically smaller vertex 
$u$ is projected by $\pi_e$). 

Given graph $G$ and constraints $\pi$, we can construct the lift of
$G$, $\widehat{G}=(V\times [k], \widehat{C})$ where:
\[
\widehat{C}_{(u,i),(v,j)} \triangleq
\begin{dcases*}
  C_{u,v} & if $\pi_{uv}(i) = j$, \\
  0 & else.
\end{dcases*}
\]
We will use $\widehat{L}$ to denote the associated Laplacian matrix of
$\widehat{G}$. Now we are ready to express Unique Games as a quadratic
programming problem:
\begin{align*}
\min_{\xint}\ & \frac12
\sum_{u<v} C_{uv} \sum_{i\in [k]}
(\xint_u(i) - \xint_v(\pi_e(i)))^2 = \frac{1}{2} \xint^T \widehat{L} \xint
, \\
\st\ \ 
 & \sum_{i\in[k]} \xint_u(i) = 1 \quad \forall u\in V,\\
& \xint\in\{0,1\}^{V\times [k]} \ , 
\end{align*} 
The corresponding SDP relaxation is given below:
\begin{align*}
\min_{x}\ & \frac12 \tr\left( \xmat^T \xmat \widehat{L}\right)\\
\st\ \   & \|\xvec_{\es}\|^2=1,\ \xmat = [\xvec_S(f)] \in \lasserrei{r}{V}{2} .
 \end{align*}

\begin{remark}
  %
  %
  Except for the problem of maximum cut, we are unable to apply
  \Cref{thm:qip} directly because there is no known way to relate the
  $r^{th}$ eigenvalue of $L$ to, say, the $\mathrm{poly}(r)^{th}$
  eigenvalue of $\widehat{L}$. We instead use the ``projection
  distance'' type bound based on column selection (similar to
  \Cref{sec:simple-anl-bisec}), after constructing an appropriate
  embedding to relate the problem to the original graph.  {\hfill
    $\Box$}
\end{remark}
\begin{remark}
  Although we do not explicitly mention in the theorem statements, we
  can provide similar guarantees in the presence of constraints
  similar to graph partitioning problems such as %
\begin{itemize}
  \itemsep=0ex
\item constraining labels available to each node,
\item constraining fraction of labels used among different subsets of
  nodes.
\end{itemize} 

For example, the guarantee for maximum cut algorithm immediately
carries over to maximum bisection with guarantees on partition sizes
similar to minimum bisection. {\hfill $\Box$}
\end{remark}

\subsubsection{Maximum cut}

We first start with the simplest problem fitting in the framework for
unique games --- finding a maximum cut in a graph.
The corresponding minimization problem is minimum uncut, where the
objective is minimizing the total cost of uncut pairs. The
corresponding $\widehat{L}$ is given by:
\[
\widehat{L}= \left(\begin{array}{cc} D & - A \\ - A^T &
    D \end{array}\right) \in \sympm^{V\times [2]}.
\]
For convenience, we also present the integral formulation below:
\begin{align*}
\min_{\xint}\ & 
\frac{1}{2}
\sum_{u<v} C_{uv}
\left[ 
(\xint_u(1) - \xint_v(0))^2
+ (\xint_u(0) - \xint_v(1))^2
\right]
, \\
\st\ \ 
& \xint_u(1)+\xint_u(0) = 1 \quad \forall u\in V,\\
& \xint\in\{0,1\}^{V\times [2]} \ .
\end{align*} 
\begin{theorem}[Maximum Cut / Minimum Uncut]
\label{thm:max-cut}
Given $G=(V,C)$, for all $\eps \in (0,1)$ and a positive integer $r$,
by solving $O(r/\eps^2)$ rounds of \lh\ relaxation, 
we can find a subset $U\subset V$ such that the
total weight of uncut edges by $U$ is at most
 \[
 \min \left\{ 1+\frac{2+\eps}{\lambda_{r+1}(\Lnorm)} , ~~
\frac{1+\eps}
{\min\left\{2 - \lambda_{n-r-1}(\Lnorm), 1\right\}} \right\} \cdot
\SDPOPT.\] 
\end{theorem}
\begin{remark}[Faster Implementation]
  \label{rem:faster-maxcut}  
  By using the faster solver from~\cite{gs12-fast}, we can implement a
  matching algorithm for~\Cref{thm:max-cut} whose running time is
  $2^{O(r/\eps^3)} n^{O(1/\eps)}$.
\end{remark}
\begin{proof}
  The first bound will follow from the more general result for Unique
  Games (\Cref{thm:ug} below), so we focus on the second bound
  claiming an approximation ratio of
  $(1+\eps)/\min\{2-\lambda_{n-r-1},1\}$.
  %
  The normalized Laplacian matrix for $\widehat{L}$ is given by
  \[
  \widehat{\Lnorm}= \left[\begin{array}{cc}
      I & -\Anorm \\ -\Anorm & I
    \end{array}\right], 
  \] where $\Anorm$ is the normalized adjacency matrix,
  $\Anorm\triangleq D^{-1/2} A D^{-1/2}$. 

  By direct substitution, it is easy to see that, for every
  eigenvector $q_i$ of constraint graph $G$'s normalized Laplacian matrix,
  $\Lnorm$, there are two corresponding eigenvectors for
  $\widehat{\Lnorm}$, $\left[\begin{array}{cc} \frac{1}{\sqrt{2}} q_i
      \\ \frac{1}{\sqrt{2}} q_i
    \end{array}\right]$ and 
  $\left[\begin{array}{cc}
      \frac{1}{\sqrt{2}} q_i \\ - \frac{1}{\sqrt{2}} q_i
    \end{array}\right]$ whose corresponding eigenvalues are given by
  $\lambda_i$ and $2-\lambda_i$ respectively. As a convention, we
  will refer to these as {\bf even} and {\bf odd} eigenvectors, respectively.
  
  For any node $u\in V$, we can express $\xvec_u(i)$ for $i\in[2]$ as 
  \[
  \xvec_u(i) = \|\xvec_u(i)\|^2 \xvec_\es + (-1)^i \|\xvec_u(1)\| \|\xvec_u(0)\| \yvec_u \ ,
  \] where $\yvec_u$ is a unit vector orthogonal to $\xvec_\es$, $\langle \xvec_\es, \yvec_u
  \rangle = 0$.
  Note that $\xvec_{\es}^\perp \xvec_u(1) = y_u = - \xvec_{\es}^\perp
  \xvec_u(0)$. Hence $\xmat^T \xvec_{\es}^{\perp} \xmat$
  has {\bf zero} correlation with even eigenvectors of $\widehat{L}$.
  Therefore we have the following identity. For any $S$:
  \begin{align*}
    \tr(\xmat^T \Pi_S^{\perp} \xmat 
    \widehat{L}) 
    = &\tr(\xmat^T \Pi_S^\perp \xmat (D + A)).
  \end{align*}
  We can slightly modify \Cref{thm:bounding-delta-r} to take into
  account only the eigenvectors of $\widehat{\Lnorm}$ with which
  $\xvec_{\es}^{\perp} \xmat$ has non-zero correlation. Using the
  bound from~\Cref{thm:bounding-delta-r}, we can then show that total
  cost of uncut edges is bounded by $(1+\eps) \frac{\SDPOPT}{
    \min\left(\lambda_{r+1}(I+\Anorm) ,1\right)}$. The proof is now
  complete by noting that $\lambda_{r+1}(I+\Anorm) = 2 -
  \lambda_{n-r-1}(\Lnorm)$.
\end{proof}

\subsubsection{Unique Games}
In this section, we prove our main result for approximating Unique
Games. For this problem, we will choose the seed set in a different
way, rather than using~\Cref{thm:bounding-delta-r}.

\begin{theorem}[Unique Games]
\label{thm:ug}
For any instance of Unique Games with label set $[k]$, constraint
graph $G=(V,C)$ and constraints $\pi$, for all $\eps \in (0,1)$ and a
positive integer $r$, the following holds: By solving $O(r/\eps)$
rounds of \lh\ relaxation, we can find a labeling $f:V\to\labs$ such that the
total cost of unsatisfied constraints is at most:
 \[
 \sum_{u<v} C_{uv} \ind{[f(v)\neq \pi_e(f(u))]} \le
 \left(1+\frac{2+\eps}{\lambda_{r+1}(\Lnorm)} \right) \SDPOPT.\]
\end{theorem}
\begin{remark}[Faster Implementation]
  \label{rem:faster-ug}  
  By using the faster solver from~\cite{gs12-fast}, we can implement a
  matching algorithm for~\Cref{thm:ug} whose running time is
  $k^{O(r/\eps)} n^{O(1)}$.
\end{remark}
\begin{proof}
  For fixed $\bestS$ (to be chosen later), we can
  use~\Cref{thm:ind-expct-over-mtx} to express the probability of
  violating constraint $\pi: \labs\to\labs$ between $u$ and $v$ as:
  \begin{align}
    \frac12
    \expcts{}{ \sum_{i,j: j=\pi(i)} (\randx_u(i) - \randx_v(j))^2 } 
    = &
    \frac12
    \sum_{i, j:j=\pi(i)} \langle \Pi_{\bestS} \xvec_u(i), \Pi_{\bestS}
    \xvec_v(j)\rangle +  
    \|\Pi_{\bestS}^\perp \xvec_u(i) \|^2 +
    \|\Pi_{\bestS}^\perp \xvec_u(\pi(i)) \|^2     \notag \\
    = & \frac12 \sum_{i, j:j=\pi(i)} \|\xvec_u(i) - \xvec_v(j)\|^2 +
    \langle \Pi_{\bestS}^\perp \xvec_u(i), \Pi_{\bestS}^\perp
    \xvec_v(j)\rangle\notag \\
    \le & \frac12 \sum_{i, j:j=\pi(i)} \|\xvec_u(i) - \xvec_v(j)\|^2 +
    \frac12 \|\Pi_{\bestS}^\perp \xvec_u(i)\|^2 + \frac12\|
    \Pi_{\bestS}^\perp \xvec_v(j)\|^2.\label{eq:ug-1246-0000}
  \end{align}
  Hence the total expected cost of violated constraints is bounded by:
  \[
  \frac12 \sum_{u<v} C_{uv} \expcts{}{ \sum_{i,j: j=\pi(i)}
    (\randx_u(i) - \randx_v(j))^2 } \le \frac12 \left( \eta + \sum_{u}
    D_u \sum_i \|\Pi_{\bestS}^\perp \xvec_u(i)\|^2 \right),
  \] where $D_u$ is the degree of node $u$, and $\eta = \tr(\xmat^T
  \xmat \widehat{L})$ is twice the SDP value.
  
  Now consider the embedding given in \Cref{thm:ug-embedding} below,
  $\{\xvec_u(i)\}_{i\in[k]}\mapsto \ymat_u$ with $\ymat =
  [\ymat_u]_u$. Then, by
  \Cref{thm:ug-embedding}~(\ref{item:ug-embed-2}):
  \[
  \sum_{u}
  D_u \sum_i \|\Pi_{\bestS}^\perp \xvec_u(i)\|^2
  \le \sum_u D_u \|\ymat_{\bestS}^\perp \ymat_u\|^2 
  = \tr(\ymat^T \ymat_{\bestS}^\perp \ymat D).
  \]  
  We can use~\Cref{thm:svd} to choose $r'=O(r/\eps)$ columns of
  $D^{1/2} \ymat$, $\bestS$. Now we can mimick the proof
  of~\Cref{lem:markov-bound} to show that:
  \[
  \tr(\ymat^T \ymat_{\bestS}^\perp \ymat D) \le \frac{(1+\eps/2)
    \tr(\ymat^T \ymat L)}{\lambda_r(\Lnorm)} \le (2+\eps)
  \frac{\tr(\xmat^T \xmat \widehat{L}) }{\lambda_r(\Lnorm)} = \frac{(2+\eps)
    \eta}{\lambda_r(\Lnorm)},
  \] where we used \Cref{thm:ug-embedding}(\ref{item:ug-embed-1})
  in the last inequality. Substituting this back
  into~\cref{eq:ug-1246-0000}, we obtain our final bound on the
  total expected cost of violated constraints:
  \[
  \frac12 \sum_{u<v} C_{uv} \expcts{}{ \sum_{i,j: j=\pi(i)}
    (\randx_u(i) - \randx_v(j))^2 } \le \opt_{\mathrm{SDP}} \Bigl( 1 +
  \frac{2+\eps}{\lambda_r(\Lnorm)} \Bigr) \ .
  \tag*{\qedhere}
  \]
\end{proof}
\subsubsection{A useful embedding}
We now turn to the embedding used in the above proof.
\begin{theorem}
  \label{thm:ug-embedding}
  Given vectors $(\xvec_{u}(i))_{u\in V, i\in \labs}$, suppose they
    satisfy the following: For any $u\in V$, whenever $f,
    g\in [k]^u$ are two different labelings of $u$, $f\neq g$,
  \[
  \langle \xvec_u(f), \xvec_u(g)\rangle = 0.
  \]  
  Then there exists an embedding $(\xvec_u(f))_{f} \mapsto \yvec_u$
  for all $u\in V$ such that:
  \begin{enumerate}
  \item \label{item:ug-embed-0}
    For any $u\in V$, $\|\yvec_u\|^2 = \sum_f \|\xvec_u(f)\|^2$.
  \item \label{item:ug-embed-1}
    For any $u, v\in V$ and any permutation $\pi\in\sym([k])$:
    \[
    \sum_{i\in [k]} \|\xvec_u(i^u) - \xvec_v({\pi(i)}^v)\|^2 
    \ge \frac{1}{2} \|\yvec_u - \yvec_v\|^2.
    \]
  \item \label{item:ug-embed-2}
    For any subset $S\subseteq V$ and any $u\in V$, if we let 
    $P_S, Q_S$ be the projection matrices onto the spans of
    $\{\xvec_s(f)\}_{s\in S, f\in[k]}$ and $\{\yvec_s\}_{s\in S}$,
    respectively, then:
    \[
    \|Q_S^\perp \yvec_u\|^2 \ge \sum_{f\in [k]^u} \|P_S^{\perp} \xvec_u(f)\|^2.
    \]
  \end{enumerate}
\end{theorem}
In the rest of this section, we will prove \Cref{thm:ug-embedding}.
Our embedding is as follows.  Assume that the vectors $\xvec_u(f)$
belong to $\R^m$. Let $e_1,e_2,\dots,e_m \in \R^m$ be the standard
basis vectors.
Define $\yvec_u \in \R^m \otimes \R^m$ as
\[ \yvec_u = \sum_{i=1}^m \sum_{f \in [k]^u} \langle
\unit{\xvec_u(f)}, e_i \rangle \xvec_u(f) \otimes e_i \ . \]

\begin{observation} For vectors $x,y \in \R^m$, $\sum_{i=1}^m \langle x, e_i\rangle\langle y,
  e_i \rangle = \langle x, y\rangle$.
\end{observation}

\noindent 
The first property of the vectors $\yvec_u$ follows from this observation easily:
\begin{align*}
  \|\yvec_u\|^2 = & \sum_i  \sum_{f,g} \langle \unit{\xvec_u(f)}, e_i\rangle\langle
  \unit{\xvec_u(g)}, e_i\rangle \langle \xvec_u(f), \xvec_u(g) \rangle  \\
  = & \sum_{f,g}  \langle \xvec_u(f), \xvec_u(g) \rangle \sum_i  \langle \unit{\xvec_u(f)}, e_i\rangle\langle
  \unit{\xvec_u(g)}, e_i \rangle  \\
  = & \sum_{f,g} \langle \xvec_u(f), \xvec_u(g)\rangle \langle \unit{\xvec_u(f)},\unit{\xvec_u(g)}\rangle \\
  = & \sum_{f} \|\xvec_u(f)\|^2. %
\end{align*}
We prove the second property in Claim \ref{ug-embed-ub-permutation-statement} 
and third one in Claim \ref{ug-embed-projection-statement} below.
\begin{claim}\label{ug-embed-ub-permutation-statement} 
  For any
  permutation $\pi\in \sym([k])$:
  \[ \frac{1}{2} \| \yvec_u- \yvec_v\|^2 \le \sum_{f\in[k]} \|\xvec_u(f) - \xvec_v({\pi(f)})\|^2 \]
\end{claim}
\begin{proof} 
  Without loss of generality, we assume $\pi$ is the
  identity permutation. We have 
  \begin{align}
    \frac{1}{2} \|\yvec_u - \yvec_v\|^2 & =
    \frac{\|\yvec_u\|^2 + \|\yvec_v\|^2}{2}
    -
    \langle \yvec_u, \yvec_v\rangle\notag \\
    &=   \frac{\|\yvec_u\|^2 + \|\yvec_v\|^2}{2} - \sum_{f,g}\langle \xvec_u(f), \xvec_v(g) \rangle \sum_i \langle \unit{\xvec_u(f)}, e_i\rangle\langle
    \unit{\xvec_v(g)}, e_i\rangle\notag \\
    & = %
    \sum_f \frac{\|\xvec_u(f)\|^2 + \|\xvec_v(f)\|^2}{2} 
    - \sum_{f,g} \langle \unit{\xvec_u(f)}, \unit{\xvec_v(g)}\rangle^{2}
    \|\xvec_u(f)\| \|\xvec_v(g)\| \notag
    \intertext{%
      The sum over all
      pairs is lower bounded by summing only the corresponding pairs:}
    & \le \frac{1}{2} 
    \sum_f
    \left(
      \|\xvec_u(f)\|^2 + \|\xvec_v(f)\|^2 - 2 \langle \xvec_u(f), \xvec_v(f)\rangle
      \langle \unit{\xvec_u(f)}, \unit{\xvec_v(f)}\rangle
    \right)\notag\\
    &= \frac{1}{2} 
    \sum_f
    \|\xvec_u(f) - \xvec_v(f)\|^2 + 
    \sum_f \langle \xvec_u(f), \xvec_v(f)\rangle
    \underbrace{
      \left(1 - \langle \unit{\xvec_u(f)}, \unit{\xvec_v(f)}\rangle
      \right)}_{\ge 0} \label{eq:ug-embed-1}
    \intertext{Since the coefficient of $\langle \xvec_u(f), \xvec_v(f)\rangle$
      is positive, we can use Cauchy-Schwarz inequality to
      replace $\langle \xvec_u(f), \xvec_v(f)\rangle$ with $\|\xvec_u(f)\|\cdot\|\xvec_v(f)\|$
      in \cref{eq:ug-embed-1} and obtain:}
    &\le \frac{1}{2} 
    \sum_f   
    \|\xvec_u(f) - \xvec_v(f)\|^2 + \sum_f\left( 
      \| \xvec_u(f) \| \cdot \| \xvec_v(f) \| - 
      \langle {\xvec_u(f)}, {\xvec_v(f)}\rangle
    \right)\label{eq:ug-embed-2}
    \intertext{Using inequality $\|\xvec_u(f)\|\cdot\|\xvec_v(f)\|
      \le \frac{1}{2} \left(\|\xvec_u(f)\|^2+\|\xvec_v(f)\|^2\right)$ on
      \cref{eq:ug-embed-2}:}
    &\le \frac{1}{2} 
    \sum_f   
    \bigg(    
    \|\xvec_u(f) - \xvec_v(f)\|^2 +
    \| \xvec_u(f) \|^2+ \| \xvec_v(f) \|^2 - 2
    \langle {\xvec_u(f)}, {\xvec_v(f)}\rangle
    \bigg)\notag\\
    & = \sum_f \|\xvec_u(f) - \xvec_v(f)\|^2. \tag*{\qedhere}
  \end{align}
\end{proof}
\begin{claim} \label{ug-embed-projection-statement}
  \[ \|Q^{\perp}_S \yvec_u\|^2 \ge \sum_f
  \|P_S^{\perp} \xvec_u(f)\|^2. \]
\end{claim}
\begin{proof}
  For any $\theta\in \R^{S}$:
  \begin{align}
    \|\yvec_u - \sum_v \theta_v \yvec_v\|^2 = ~\sum_{i=1}^m \bigg\|\sum_f
    \langle \unit{\xvec_u(f)},e_i\rangle \xvec_u(f)- \underbrace{\sum_{v\in
        S,g} \theta_v \langle \unit{\xvec_v(g)},e_i\rangle \xvec_v(g) }_{P_S
      \Theta_i}\bigg\|^2.
    \label{eq:ug-embed-3}
  \end{align}
  Substituting $\alpha_f = P_S^{\perp} \xvec_u(f)$ and $\beta_f = P_S
  \xvec_u(f)$ in \cref{eq:ug-embed-3}:
  \begin{align*}
    = & ~ \sum_{i=1}^m \bigg\|\sum_f \langle \unit{\xvec_u(f)},e_i\rangle (\alpha_f + \beta_f)- P_S \Theta_i\bigg\|^2 \\
    = & ~\sum_{i=1}^m \bigg\|\sum_f \langle
    \unit{\xvec_u(f)},e_i\rangle \alpha_f\bigg\|^2
    +\bigg\|\sum_f \langle \unit{\xvec_u(f)},e_i\rangle \beta_f- P_S \Theta_i\bigg\|^2  \\
    \ge &\sum_{i=1}^m \bigg\|\sum_f \langle\unit{\xvec_u(f)},e_i\rangle \alpha_f\bigg\|^2 \\
    = & ~  \sum_{f,g} \langle \alpha_f, \alpha_g\rangle \sum_{i=1}^m \langle \unit{\xvec_u(f)},e_i\rangle \langle \unit{\xvec_u(g)},e_i\rangle \\
    = & ~ \sum_{f,g} \langle \alpha_f, \alpha_g\rangle \langle
    \unit{\xvec_u(f)}, \unit{\xvec_u(g)}\rangle = \sum_f
    \|\alpha_f\|^2 = \sum_f \|P_S^{\perp}
    \xvec_u(f)\|^2. \tag*{\qedhere}
  \end{align*} 
\end{proof}    
This concludes the proof of \Cref{thm:ug-embedding}, therefore also
the proof of \Cref{thm:ug}.

\subsection{Independent Set}
\label{sec:is-gen}
\def\dmax{\mathrm{d}_{\max}}
Our final algorithmic result is on finding independent sets in a
graph. For simplicity, we focus on unweighted graphs though the
extension for graphs with non-negative vertex weights is
straightforward. As usual, we denote by $\alpha(G)$ the size of the
largest independent set in $G$.  Let $\dmax$ denote the maximum degree
of a vertex of $G$.
\begin{theorem}\label{thm:is}
  Given $0<\eps<1$, positive integer $r$, a graph $G$ with $\dmax \ge
  3$, by solving $O(r/\eps^2)$ rounds of \lh\ relaxation, we can find
  an independent set $I\subseteq V$ such that
  \begin{equation}
    \label{eq:is-size}
    |I| \ge  \alpha(G) \cdot \min\left\{
      \frac{1}{2 \dmax}
      \left(\frac{1}{(1-\eps)\min\{2-\lambda_{n-r-1}(\Lnorm),1\}}
        -1\right)^{-1}, 1\right\}.
  \end{equation}
\end{theorem}
\begin{remark}[Faster Implementation]
  \label{rem:faster-is}  
  By using the faster solver from~\cite{gs12-fast}, we can implement a
  matching algorithm for~\Cref{thm:is} whose running time is $2^{O(r)}
  n^{O(1)}$.
\end{remark}
\begin{remark}
  The above bound (\ref{eq:is-size}) implies that if
  $\lambda_{n-r-1}$, which is the $(r+1)^{\mathrm{st}}$ largest
  eigenvalue of the normalized Laplacian $\Lnorm$, is very close to
  $1$, then we can find large independent sets in $n^{O(r/\eps^2)}$
  time. In particular, if it is at most $1+\frac{1}{4\dmax}$, then
  taking $\eps =O(1/\dmax)$, we can find an optimal independent
  set. The best approximation ratio for independent set in terms of
  $\dmax$ is about $O\left(\frac{\dmax \cdot \log \log \dmax}{\log
      \dmax}\right)$~\cite{Halldorsson98,halperin02}. The bound in
  \cref{eq:is-size} gives a better approximation ratio when
  $\lambda_{n-r-1} \leq 1 + O\left( \frac{1}{\log \dmax}\right)$.
  {\hfill $\Box$}
\end{remark}
\begin{proof}{\it (of Theorem~\ref{thm:is})}
  Consider the following integer program for finding largest
  independent set in $G$:
  \begin{align*}
    \max\ & \sum_u \xint_u\\
    \st\ \ &
    \xint_u \xint_v = 0 \qquad\mbox{for any edge $e=(u,v)\in E$ ,}\notag
    \\
    & \xint\in \{0,1\}^{V}.\notag
  \end{align*} 
  Its SDP relaxation is given by:
  \begin{align*}
    \max\ & \sum_u \|\xvec_u\|^2\\
    \st\ \ &
    \langle \xvec_u, \xvec_v\rangle = 0
    \qquad\mbox{for any edge $e=(u,v)\in E$ ,}\notag
    \\
    & \|\xvec_{\es}\|^2=1,\ 
    \xmat = [\xvec_S(f)] \in \lasserrei{r}{V}{2}; \notag
  \end{align*} 
  with value $\sum_{u} \|\xvec_u\|^2 = \mu$. 

  We will choose $\bestS$ by applying~\Cref{thm:bounding-delta-r} on
  matrices $I_V+\Anorm$ and $I_V$, so that:
  \begin{eqnarray*}
    \tr(\xmat^T \Pi_{\bestS}^{\perp} \xmat I_V) + 
    \tr(\xmat^T \Pi_{\bestS}^{} \xmat ( I_V+\Anorm)) & 
    \le &  \frac{\tr(\xmat^T \xmat ( I_V+\Anorm))}{\lambda'} \\
    &  = & \frac{1}{\lambda'} \tr(\xmat^T \xmat I_V) 
    \quad \mbox{since} \quad \tr(X^T X \Anorm)=0,
  \end{eqnarray*}
  where $\lambda' =  \frac{\min\{\lambda_{r+1}(I+\Anorm), 1\}}{1+\eps} =
  \frac{\min\{2-\lambda_{n-r-1}(\Lnorm), 1\}}{1+\eps}$.

  After sampling $f$ and $\randx\in \{0,1\}^V$, we convert
  $\randx$ into an independent set as follows.
  \begin{enumerate}
  \item For each $u$, if $\randx_u=1$ then let $I \gets I \cup \{u\}$
    with probability $p_u$ which we will specify later.
  \item After the first step, for each edge $e=\{u,v\}$, if $\{u,v\}\subseteq I$, we 
    choose one end point randomly, say $u$, and set $I\gets I\setminus \{u\}$.
  \end{enumerate}
  By construction, final $I$ is an independent set.
  
  Note that for any $u$, the probability that $u$ will be in the final
  independent set $I$ is at least:
  \begin{align}
    \probs{}{u\in I} \ge &
    \expcts{}{p_u \randx_u} -\frac{1}{2}
    \expcts{}{ \sum_{v\in N(u)} p_u p_v \randx_u\randx_v } \nonumber \\
    = & p_u \|\xvec_u\|^2 -\frac{1}{2}
    \sum_{v\in N(u)} p_u p_v \langle \Pi_{\bestS} \xvec_u, \Pi_{\bestS} \xvec_v\rangle. \label{eq:hwe}
  \end{align}
  By~\cref{eq:hwe}, the expected size of the independent set found by
  the algorithm satisfies
  \begin{equation}
    \label{eq:ind-set-size}
    \expcts{}{|I|} \ge  \sum_u p_u \|\xvec_u\|^2 
    -\sum_{\{u,v\} \in E} p_u p_v \langle \Pi_{\bestS} \xvec_u, \Pi_{\bestS} \xvec_v\rangle.  
  \end{equation} 
  Note that for every edge $\{u,v\} \in E$, 
  \begin{equation}
    \label{eq:vfr-1}
    \langle \Pi_{\bestS} x_u, \Pi_{\bestS} x_v\rangle = 
    \sum_{f\in [2]^S} \frac{\langle \xvec_{\bestS}(f), x_u\rangle \langle \xvec_{\bestS}(f), \xvec_v\rangle}{\|\xvec_{\bestS}(f)\|^2} \ge 0 \ . 
  \end{equation}
  We now consider two cases. 
  
  \medskip
  \noindent {\sf Case 1:} $\langle \Pi_{\bestS} \xvec_u, \Pi_{\bestS} \xvec_v \rangle = 0$
  for all edges $\{u,v\} \in E$. In this case, we take $p_u=1$ for all
  $u \in V$, and by~\cref{eq:ind-set-size}, we find an independent set
  of expected size at least $\mu \ge \alpha(G)$.
  
  \medskip
  \noindent {\sf Case 2:} In this case, we have
  \begin{equation}
    \label{eq:vfr-2}
    \sum_{\{u,v\} \in E} \langle \Pi_{\bestS} \xvec_u, \Pi_{\bestS} \xvec_v\rangle 
    = \frac{1}{2} \tr(\xmat^T \Pi_{\bestS} \xmat A) > 0 \ , 
  \end{equation}
  where $A$ is the adjacency matrix of $G$.  Let $\Anorm = D^{-1/2} A
  D^{-1/2}$ be the normalized adjacency matrix, and define
  \begin{equation}
    \xi \triangleq \frac{\tr(\xmat^T \Pi_{\bestS} \xmat \Anorm)}
    {\tr(\xmat^T \xmat I_V)}.\label{eq:is-xi-def}
  \end{equation}
  By \cref{eq:vfr-1,eq:vfr-2}, we have $\xi > 0$.
  
  We now pick $p_u = \frac{\alpha}{\sqrt{d_u}}$ for all $u\in V$,
  where we will optimize the choice of $\alpha$ shortly.  For this
  choice, we have
  \begin{eqnarray*}
    \expcts{}{|I|} & \ge & \alpha \sum_u \frac{1}{\sqrt{d_u}} \|\xvec_u\|^2 
    -\frac{1}{2} \alpha^2 \tr(\xmat^T \Pi_{\bestS} \xmat \Anorm)\\
    & \ge & \frac{\alpha}{\sqrt{\dmax}} \sum_u \|\xvec_u\|^2 
    -\frac{1}{2} \alpha^2 \tr(\xmat^T \Pi_{\bestS} \xmat \Anorm)\\
    & = & \mu \bigg(
    \frac{\alpha}{\sqrt{\dmax}} 
    -\frac{1}{2} \alpha^2 
    \underbrace{\frac{\tr(\xmat^T \Pi_{\bestS} \xmat \Anorm)}
      {\tr(\xmat^T \xmat I_V)}}_{\xi}
    \bigg)
  \end{eqnarray*}
  This expression is maximized when $\alpha = \frac{1}{\xi \cdot
    \sqrt{\dmax}}$, for which it becomes:
  \begin{equation}
    \expcts{}{|I|} 
    \ge 
    \frac{\mu}{2 \dmax} \frac{1}{\xi}.\label{eq:eye-size}
\end{equation}
We know that:
\begin{align*}
  \frac{\tr(\xmat^T \Pi_{\bestS}^{\perp} \xmat) + 
    \tr(\xmat^T \Pi_{\bestS}^{} \xmat (I_v+\Anorm))}{\tr(\xmat^T \xmat I_V)} 
  = & \frac{\tr(\xmat^T \Pi_{\bestS}^{\perp} \xmat I_V) + 
    \tr(\xmat^T \Pi_{\bestS}^{} \xmat I_V )+
    \tr(\xmat^T \Pi_{\bestS}^{} \xmat \Anorm)
  }{\tr(\xmat^T \xmat I_V)} \\
  = & \frac{\tr(\xmat^T \xmat I_V) + 
    \tr(\xmat^T \Pi_{\bestS}^{} \xmat \Anorm)
  }{\tr(\xmat^T \xmat I_V)}  = 1 + \xi.
\end{align*}
Thus, for our choice of $\bestS$, we obtain that $ \xi \le
\frac{1}{\lambda'}-1$. Substituting this back into
\cref{eq:eye-size}, we have
\[
\expcts{}{|I|} \ge \frac{ \mu }{2 \dmax} \frac{1}{1/\lambda' - 1}.
\tag*{\qedhere} \]

\end{proof}

\section{Algorithms Based on Threshold Based Rounding}
\label{sec:app-threshold}
In this section, we will present approximation algorithms which use the
threshold based rounding scheme. All problems share the following
structure: Given a capacity graph, $G=(V,C)$, and a \emph{connected} demand
graph, $H=(V,D)$, find  a non-empty $U \subset V$ which minimizes the
total capacity cut subject to various constraints involving the total
demand cut.

Recall that, for $L_H$ and $\Gamma_H$ being the Laplacian and
edge-node incidence matrices of graph $H$, respectively, we have $L_H
= \Gamma_H^T \Gamma_H$. Since $H$ is connected, null space of $L_H$
consists only of constant vector. Hence $\nll(L_H) \subseteq
\nll(L_G)$. Also the width of $\Gamma_H$ is $2$. Therefore matrices
$L_G$ and $L_H$ satisfy all the requirements
in~\Cref{lem:markov-bound}, which we use for seed selection. The
following is the randomized rounding algorithm we will use in this
section:
\begin{enumerate}
\itemsep=0.5ex
\item Choose seeds $\bestS$ as described in~\Cref{lem:markov-bound}
  where $A \triangleq L_G$, $B \triangleq L_H$ and $\Gamma \triangleq
  \Gamma_H$.
\item \label{item:sample-f} Sample $f:\bestS \to \labs$ with
  probability $\|\xvec(f)\|^2$.
\item \label{item:sample-tau}
  Choose a threshold $\tau \in (0,1]$ uniformly at random.
\item For each $u\in V$, set $\xint_u \gets 1$ if $\|\xvec_{u\mid
    f}\|^2 \ge \tau$ and $\xint_u \gets 0$ otherwise.
\end{enumerate}
In all our analysis, we will prove that the procedure will output
correct answer with positive probability. But we can easily
derandomize the above procedure by enumerating over all:
\begin{enumerate}[(i)]
\itemsep=0.5ex
\item $f:\bestS \to \{0,1\}$ in step~(\ref{item:sample-f}),
\item $\tau \in \{ \|\xvec_{u\mid f}\|^2 \mid u \in V\}$ in
  step~(\ref{item:sample-tau}).
\end{enumerate}
After these changes, the running time of rounding is $2^{O(r)}
\poly(n)$, which does not affect the total running time.

\subsection{Sparsest Cut and Variations}
\label{sec:sc}
In the problem of~\nusc, our goal is to minimize the ratio of total
capacity cut versus total demand cut:
\begin{equation}
  \label{eq:nusc-bin-form}
  \Phi(G,H) \triangleq
  \min_{\xint\in \{0,1\}^V:  L_H \xint\neq 0 } \frac{\xint^T L_G \xint}{\xint^T L_H \xint}.
\end{equation}
This includes problems such as: Uniform Sparsest Cut, Normalized Cut,
etc... (see also~\Cref{fig:variations-of-sc}).
\begin{figure}[bt]
  \centering
  \begin{tabular}{|l|c|c|}
    \hline
    {\bf Problem} & {\bf Denominator} & 
    {\bf $\lambda_r(G,H)$ Becomes }  \\
    \hline
    Uniform Sparsest Cut & 
    $\frac{1}{n} \sum_u \xint_u ( n - \sum_u \xint_u )$
    &
    $\lambda_r(L)$ \\
    \hline
    Edge Expansion & 
    $\min( \sum_u \xint_u , n - \sum_u \xint_u )$&
    $\lambda_r(L)$ \\
    \hline
    Normalized Cut &
    $\frac{1}{m} \sum_u d_u \xint_u ( m - \sum_u d_u \xint_u )$&
    $\lambda_r(\Lnorm)$ \\
    \hline
    Conductance & 
    $\min(\sum_u d_u \xint_u , m - \sum_u d_u \xint_u )$&
    $\lambda_r(\Lnorm)$ \\
    \hline
  \end{tabular}
  \caption{Some variations of Sparsest Cut problem
    with formulations similar to~\cref{eq:nusc-bin-form}.
    In all cases,
    \Cref{cor:final-bnd},
    \Cref{thm:nusc-int-gap,cor:final-bnd-regimes} hold.  In the above
    table, $d$ denotes the degrees and $m$ denotes the total capacity.
  }
  \label{fig:variations-of-sc}
\end{figure}
The following is a relaxation of~\cref{eq:nusc-bin-form}:
\begin{align}
\label{eq:sc-sdp0}
\min \quad& \dfrac{\sum_{u<v} C_{u,v} \| \xvec_u - \xvec_v
  \|^2}{\sum_{u<v} D_{u,v}
  \| \xvec_u - \xvec_v \|^2} = \dfrac{\tr(\xmat^T \xmat
  L_G)}{\tr(\xmat^T \xmat L_H)} \\
\mathrm{st}\quad & \sum_{u<v} D_{u,v}
\| \xvec_u - \xvec_v \|^2 > 0, \notag\\
    & \|\xvec_{\es}\|^2=1,\ 
    \xmat = [\xvec_S(f)] \in \lasserrei{r}{V}{2}. \notag
\end{align} 
We will refer to the optimum of \cref{eq:sc-sdp0} as $\Phi_{SDP}$.
Even though~\cref{eq:sc-sdp0} is not convex (it is quasi-convex),
there is an equivalent convex formulation.
\begin{lemma} \label{lem:sdp-eq}
The following SDP is equivalent to~\cref{eq:sc-sdp0}:
\begin{align}
\label{eq:sc-sdp}
\min \quad& \tr(\wmat^T \wmat L_G)
\\ %
\mathrm{st}\quad & \tr(\wmat^T \wmat L_H) = 1, \notag \\
    & \|\wvec_{\es}\|^2>0,\ 
    \wmat = [\wvec_S(f)] \in \lasserrei{r}{V}{2}; \notag
\end{align} 
\end{lemma}
\begin{remark}
  The constraint $\|\wvec_{\es}\|^2 > 0$ in~\cref{eq:sc-sdp} is
  redundant, but we included it for the sake of clarity.
\end{remark}
\begin{proof} [(of~\Cref{lem:sdp-eq})] Given a feasible solution
  $\xmat$ for formulation \eqref{eq:sc-sdp0}, let $\wmat =
  [\wvec_S(f)]$ be $\wmat \gets \frac{1}{\sqrt{\tr(\xmat^T \xmat
      L_H)}} \xmat$.  It is easy to see that $\sum_{u<v} D_{u,v} \|
  \wvec_u - \wvec_v \|^2=1$ and objective values are equal.
  Finally $\|\wvec_{\es}\|^2 = \frac{1} {\sum_{u<v} D_{u,v} \| \xvec_u
    - \xvec_v \|^2} > 0$ since $0<\sum_{u<v} D_{u,v} \| \xvec_u -
  \xvec_v \|^2.$

  For the other direction of equivalence, suppose $\wvec$ is a
  feasible solution of~\cref{eq:sc-sdp}. For each $T\in \binom{V}{\le
    r'}, f\in \{0,1\}^T$, let $\xvec_T(f) \gets
  \frac{1}{\|\wvec_{\es}\|} \wvec_T(f)$. It is easy to see that the
  objective values are equal. The rest of the proof for $\xvec$ being
  a feasible solution of~\cref{eq:sc-sdp0} follows similarly to the
  previous direction.
\end{proof}
\begin{remark}
  Our rounding procedure is scale invariant, thus~\cref{eq:sc-sdp} is
  sufficient for rounding purposes. But we chose to first
  present~\cref{eq:sc-sdp0} as it is more intuitive.
\end{remark}
\newcommand{\scsdp}{\Phi^{\mathrm{SDP}}}
\begin{theorem}
  \label{cor:final-bnd}
  Given capacity graph $G=(V,C)$ and a connected demand graph
  $H=(V,D)$, positive $r$ and $\eps$; one can find $\randx \in
  \{0,1\}^V$ such that
  \[
  \frac{\randx^T L_G \randx}{\randx^T L_H \randx} 
  \le \Phi_{SDP} \left( 1 - (1+\eps) \frac{\Phi_{SDP}}{\lambda_{r}(G,H)} \right)^{-1}
  \] provided that $\lambda_r(G,H) \ge (1+\eps) \Phi_{SDP}$, by rounding
  $O(r/\eps)$-rounds of \lh\ relaxation.
\end{theorem}
\begin{remark}[Faster Implementation]
  \label{rem:faster-sc}  
  By using the faster solver from~\cite{gs12-fast}, we can implement a
  matching algorithm for~\Cref{cor:final-bnd} whose running time is
  $2^{O(r/\eps)} n^{O(1)}$.
\end{remark}
\begin{proof} For fixed $f$, the probability of separating $u$ and $v$
  is at most $\|\xvec_{u\mid f} - \xvec_{v\mid f}\|^2$. Taking
  expectation over $f$, we see that:
  \[
  \expcts{f,\tau}{(\randx_u - \randx_v)^2} \le 
  \expcts{f}{
  \|\xvec_{u\mid f} - \xvec_{v\mid f}\|^2} = \|\xvec_u - \xvec_v\|^2.
  \] 
  By~\Cref{thm:threshold-lb}, we can lower bound the probability of
  separating $u$ and $v$ by:
  \[
  \expcts{f,\tau}{(\randx_u - \randx_v)^2} \ge 
  \|\xvec_u - \xvec_v\|^2 - \|\Pi_{\bestS}^\perp(\xvec_u - \xvec_v)\|^2.
  \] 
  Consequently,
  \[
  \frac{\expcts{}{\randx^T L_G \randx}} {\expcts{}{\randx^T L_H
      \randx}} \le \frac{ \tr(\xmat^T \xmat L_G) }{ \tr(\xmat^T \xmat
    L_H) - \tr(\xmat^T \Pi_{\bestS}^\perp \xmat L_H) }.
  \]  
  Since we choose $\bestS$ as in~\Cref{lem:markov-bound}, we can
  use~\cref{eq:delta-r:markov-bound-0} to bound $\tr(\xmat^T
  \Pi_{\bestS}^\perp \xmat L_H)$:
  \[
  \tr(\xmat^T \Pi_{\bestS}^\perp \xmat L_H) \le (1+\eps)
  \frac{\tr(\xmat^T \xmat L_G)}{\lambda_r}
  = \frac{\tr(\xmat^T \xmat L_H)}{\lambda_r} \Phi_{SDP}.
  \] Substituting this back into the previous upper bound yields:
  \[
  \frac{\expcts{f,\tau}{\randx^T L_G \randx}} {\expcts{}{\randx^T L_H
      \randx}} \le \frac{ \tr(\xmat^T \xmat L_G) }{ \tr(\xmat^T \xmat
    L_H) (1 - (1+\eps) \frac{\Phi_{SDP}}{\lambda_r})}
  = \frac{\Phi_{SDP}}{1 - (1+\eps) \frac{\Phi_{SDP}}{\lambda_r}}.
  \]
  This implies the existence of $f$ and $\tau$ for which the claim holds.  
\end{proof}
Even though~\Cref{cor:final-bnd} is not guaranteed to output a good
solution when $\lambda_r \le \Phi_{SDP}$, we can still obtain an
unconditional bound on the integrality gap:
\begin{corollary}
  \label{thm:nusc-int-gap}
  The integrality gap after $O(r)$ rounds is bounded by:
  \[
  \frac{\Phi}{\Phi_{SDP}}\le
  O \Big[ \max\big( \Phi / \lambda_r , 1\big) \Big].
  \]
\end{corollary}
\begin{proof}
  Let $\eps \gets \frac{1}{2}$. If $\frac{\lambda}{\Phi_{SDP}}\ge
  \frac{1+\eps}{1-\eps} = 3$; then the algorithm will output a subset
  whose sparsity is at most $2 \Phi_{SDP}$, which implies
  $\frac{\Phi}{\Phi_{SDP}} \le 2$. Otherwise, $\frac{\lambda}{3} <
  \Phi_{SDP}$. Taking the reciprocal of both sides and multiplying by
  $\Phi$, we see that $\frac{\Phi}{\Phi_{SDP}} < \frac{3
    \Phi}{\lambda}.$
\end{proof}
Finally we highlight the two interesting regimes
of~\Cref{cor:final-bnd} in~\Cref{cor:final-bnd-regimes}.
\begin{corollary}
  \label{cor:final-bnd-regimes}
  Given capacity graph $G=(V,C)$ and a connected demand graph $H=(V,D)$,
  positive $r$ and $\delta$; one can find a subset whose 
  sparsity is at most:
  \begin{itemize}
  \item (Near Optimal) $\Phi (1+\delta)$ if $\Phi < \frac{1}{2} \delta \cdot
    \lambda_{r}$; %
  \item (Constant Factor) $\frac{\Phi}{\delta}$ if $\Phi < (1-2 \delta)
    \lambda_{r}$. %
  \end{itemize}
  by rounding $O(r/\delta)$-rounds of \lh\ relaxation.
\end{corollary}
\subsection{Minimizing Capacity Cut Under Packing Constraints}
\label{sec:qcut}
In this section, we will consider the problem of finding a subset of
nodes with respect to some packing constraint while minimizing the
total capacity cut. This includes problems such as balanced separator,
minimum bisection. 
Formally, given a graph $G$ with Laplacian matrix
$L$, a non-negative diagonal matrix $D = \diag(d) \in \symm^V_+$
(think of $d\in \R^V_+$ as node weights with $\sum_u d_u = m$) and
positive real $\mu$; we consider the following problem:
\begin{equation}
  \label{eq:balanced-cut}
  \begin{array}{rcl}
    \min&& {\xint^T L \xint} \\ %
    \st&& \sum_u d_u \xint_u = \xint^T D \xint = \mu,
    \\
    && \xint \in \{0,1\}^V.
  \end{array} 
\end{equation}
Our main result is the following.
\def\wmu{\widetilde{\mu}}
\begin{theorem}
  \label{thm:qcut-approx-guarantee}
  Given a problem of the form \cref{eq:balanced-cut} and positive
  integer $r$, if $\delta>0$ satisfies
  \[
  \lambda_r(L,D) \ge \frac{1+\epsilon}{\delta} \frac{\eta}{\mu^2/m},
  \] then, by rounding $O(r/\eps)$ rounds of \lh\ relaxation, one can
  find $\randx \in \{0,1\}^V$ such that:
  \begin{enumerate}[(i)]
  \item $\big|\randx^T D \randx - \mu\big| \le \delta^{1/3} \mu$,
  \item $\randx^T L \randx \le \dfrac{\eta}{1-\delta^{1/3}}$.
  \end{enumerate} 
  Here $\eta$ is the optimum value of relaxation.
\end{theorem}
\begin{remark}[Faster Implementation]
  \label{rem:faster-qcut}  
  By using the faster solver from~\cite{gs12-fast}, we can implement a
  matching algorithm for~\Cref{thm:qcut-approx-guarantee} whose
  running time is $2^{O(r/\eps)} n^{O(1)}$.
\end{remark}
\begin{proof} %
  Since we choose $\bestS$ as in~\Cref{lem:markov-bound}, we can
  use~\cref{eq:delta-r:markov-bound-0} to bound $\tr(\xmat^T
  \Pi_{\bestS}^\perp \xmat D)$:
  \[
  \tr(\xmat^T \Pi_{\bestS}^\perp \xmat D) \le (1+\eps)
  \frac{\tr(\xmat^T \xmat L)}{\lambda_r} \le \frac{\delta \mu^2}{m}.
  \]
  Recall that $\tr(\xmat^T \Pi_{\bestS}^\perp \xmat D) =
  \expcts{f}{\sum_u d_u \|\xvec_{\es\mid f}^\perp \xvec_{u\mid
      f}\|^2}$. Hence there exists $f:S\to\{0,1\}$ for which:
\begin{align}
  \label{eq:qcut-delta-defn}
  \frac{\delta \mu^2 }{m} \ge 
  \sum_{u} d_u \|\xvec_{\es\mid f}^\perp 
    \xvec_{u\mid f} \|^2.
\end{align}
We define $y_u\triangleq \|\xvec_{u\mid f}\|^2$ so that
$\|\xvec_{\es\mid f}^\perp \xvec_{u\mid f} \|^2 = \|\xvec_{u\mid
  f}\|^2 - \|\xvec_{u\mid f}\|^4 = y_u (1-y_u)$ and $\delta
\frac{\mu^2}{m} \ge \sum_u {d_u} y_u(1-y_u) $. 
\begin{claim}
  \label{thm:qcut-var-weight}
  $ \vars{\tau}{ \sum_u { d_u}{\randx_u} } \le\delta \mu^2$. In
  particular,
  \[\probs{\tau}{\big| \sum_u { d_u}{\randx_u}-\mu\big| > \delta^{1/3} \mu} 
  < \delta^{1/3}.\]
\end{claim}
\begin{proof}
  We can express $\vars{\tau}{ \sum_u { d_u}{\randx_u} }$ as:
  \begin{align*}
    \vars{\tau}{ \sum_u { d_u}{\randx_u} } = & \sum_{u,v} { d_u d_v}
    \expcts{}{\randx_u \randx_v} - \left[\sum_u d_u \expcts{}{\randx_u}\right]^2
    = \sum_{u,v} d_u d_v {\min(y_u, y_v)} - \mu^2  \\
    = &\frac{1}{2} \sum_{u,v} d_u d_v (y_u + y_v)
    - \frac{1}{2} \sum_{u,v} d_u d_v {|y_u - y_v|}- \mu^2 \\
    \le & m \sum d_u {y_u}
    - \frac{1}{2} \sum_{u,v} d_u d_v {(y_u - y_v)^2}- \mu^2 \\
    = & m \sum_{u} d_u {y_u (1-y_u)}
    + \frac{1}{2} \sum_{u,v} d_u d_v (2 y_u y_v)- \mu^2 \\
    = & m \sum_{u}d_u {y_u (1-y_u)} \le \delta \mu^2. 
  \end{align*}
  The second bound follows from the first one via Chebyshev's inequality.
\end{proof}
Given~\Cref{thm:qcut-var-weight}, the rest of the proof is easy.
For random $\tau$, with probability $> 1 - \delta^{1/3}$, the
corresponding $\randx$ satisfies the first condition as proven
in~\Cref{thm:qcut-var-weight}. Since the expected cut cost is bounded
by $\eta$ in~\Cref{thm:threshold-ub}, we can use Markov's inequality
to show that $\randx^T L \randx$ will be at most $(1-\delta^{1/3})^{-1}
\eta$ with probability $\le 1 - \delta^{1/3}$. Hence with non-negative
probability, both properties are satisfied simultaneously. {\hfill $\Box$}
\end{proof}

\section*{Acknowledgments}
We thank Sanjeev Arora for useful comments on an earlier manuscript.
\bibliographystyle{alpha}
\bibliography{references}

\appendix
\section{Analysis of Other Rounding Algorithms Using Column Selection
  Framework}
\label{sec:other-rounding-col-sel}
\subsection{$2$-CSPs}
\label{sec:csp}
Given a $2$-CSP problem on variables $[n]$ and labels $\labs$, let $G$
be its constraint graph. For convenience, we assume $G$ is regular;
however all our bounds still hold when $G$ is non-regular.  We use $A$
to denote $G$'s normalized adjacency matrix and $\lambda_i$ to denote
the $i^{th}$ smallest eigenvalue of $G$'s normalized Laplacian matrix.
Finally we will use $uv \sim G$ to denote sampling a constraint with
probability proportional to the weight of constraint between $u$ and
$v$.

\myparagraph{Embedding} 
Consider the embedding used in Lemma 5.3 of~\cite{brs11} which is used
to convert $k$ vectors $\xvec_u(i)$ into a single vector.  Given a
partial assignment $f \in \labs^S$ and $u \in [n]$ with $(\xvec_{u\mid
  f}(i))_{i\in \labs} \subset \R^{[m]}$, we define $\xmat_u(f)$ as the
following vector.
\begin{equation} \label{eq:brs-embed} \xmat_u(f) \triangleq
  \frac{1}{\sqrt{k}} \sum_{j} \frac{(\xvec_{\es\mid f}^{\perp}
    \xvec_{u\mid f}(j))^{\otimes 2}}{\|\xvec_{\es\mid f}^{\perp}
    \xvec_{u\mid f}(j)\|}
\end{equation}
\myparagraph{Seed Selection and Rounding} 
We will give only an overview of the seed selection procedure.
At $i^{th}$ level, we choose a seed set of size $O(r/\eps)$, $S_i$,
from the matrix $\xmat(f_i) = [\xmat_u(f_i)]_{u\in [n]}$ where
$\xmat_u$'s are defined in~\cref{eq:brs-embed}. After choosing seed
set $S_i$, we sample an assignment $g_i\in \labs^{S_i}$ (conditioned
on $f_{i-1}$) that satisfies
\[
\delta_{f_{i-1},g_i} \le \mathbb{E}_{g \sim \|\xvec_{S\mid
    f_{i-1}}(g)\|^2} \left[ \delta_{f_{i-1},g} \right] \] where
$\delta_f$ is defined in~\cref{eq:brs-delta} and set $f_i \gets
f_{i-1}\circ g_i$. We repeat the seed selection procedure as long as
$\epsilon_{f_i} > \eps$ where $\epsilon_{f}$ is as defined in
\cref{eq:brs-err}.

The rounding procedure remains the same --- independent labeling for
each CSP variable from the respective conditional
distributions. Formally, for each variable $u \in [n]$, we choose a
label $i\in \labs$ with probability $\|\xvec_{u\mid f_i}(i)\|^2$
independently at random.  In~\Cref{thm:brs-final}, we will show that
$\ell = {O}\left( \frac{k^2}{\eps^2} \right)$, i.e. seed selection
will terminate after choosing at most $\ell$ sets.

\myparagraph{Analysis}
Let us begin by defining the quantity
\begin{align}
  \epsilon_{f} \triangleq&
  \mathbb{E}_{uv\sim G} \sum_{(i,j) \in [k]^2}%
  \left| \mathbb{E}\left[ \mathcal{X}_{uv\mid f}(ij) \right] -
    \mathbb{E}\left[ \mathcal{X}_{u\mid f}(i) \right]
    \mathbb{E}\left[ \mathcal{X}_{v\mid f}(j) \right]\right| \notag\\
  = & \mathbb{E}_{uv\sim G} \sum_{i,j}%
  \left| \mathrm{Cov}\left[\mathcal{X}_{u\mid f}(i),
      \mathcal{X}_{v\mid f}(j) \right]\right| %
  = \mathbb{E}_{uv\sim G} \sum_{i,j}%
  \left| \langle \xvec_{\es\mid f}^{\perp} \xvec_{u\mid f}(i),
    \xvec_{\es\mid f}^{\perp} \xvec_{u\mid f}(j) \rangle\right|
  .\label{eq:brs-err}
\end{align}
As shown in \cite{brs11}, the above gives an upper bound on the
expected extra fraction of unsatisfied constraints in the rounded
solution compared to the SDP optimum (when performing rounding after
conditioning on assignment $f$). Therefore, when $\eps_f \le \eps$, we
get an additive $\eps$-error approximation. Our goal is prove (which
we will do in \Cref{thm:brs-final}) that for $\ell \le
\tilde{O}(k/\eps)$, we must have $\eps_{f_\ell} \le
\eps$.  %

\noindent If we define the quantity $\delta_f$ measuring the expected
total variances of each $\mathcal{X}_{u|f}(i)$ as
\begin{equation}\label{eq:brs-delta}
  \delta_f \triangleq 
  \mathbb{E}_u \sum_{i\in [k]} \mathrm{Var}\left[\mathcal{X}_{u\mid f}(i)\right]
  =
  \mathbb{E}_u \sum_{i \in [k]} \|\xvec_{\es\mid f}^{\perp} \xvec_{u\mid f}(i)\|^2 \ ,
\end{equation}
then it is easy to see that $\epsilon_f \le k \delta_f$ by Cauchy-Schwarz. 

\noindent
We will first relate \cref{eq:brs-err} to the inner products of the
embedded vectors $\xmat_u(f)$.
\begin{claim}%
 \label{clm:brs-lb}
 $ \mathbb{E}_{uv\sim G}\left[ \langle \xmat_u(f), \xmat_v(f)\rangle
 \right] \ge \left( \frac{\epsilon_{f}}{k}\right)^2$.
\end{claim}
\begin{proof}
  We have 
  \begin{equation}
    \label{eq:ali09}
    k \langle \xmat_u(f), \xmat_v(f)\rangle 
    = \sum_{ij} \frac{\langle \xvec_{\es\mid f}^{\perp} \xvec_{u\mid f}(i),
      \xvec_{\es\mid f}^{\perp} \xvec_{v\mid f}(j)
      \rangle^2 }{\|\xvec_{\es\mid f}^{\perp} \xvec_{u\mid f}(i)\|
      \|\xvec_{\es\mid f}^{\perp} \xvec_{v\mid f}(j) \|}
    \ge \frac{ \left(
        \sum_{ij} \left|\langle \xvec_{\es\mid f}^{\perp} \xvec_{u\mid f}(i),
          \xvec_{\es\mid f}^{\perp} \xvec_{v\mid f}(j)
          \rangle\right|\right)^2}{\sum_{ij} \|\xvec_{\es\mid f}^{\perp} \xvec_{u\mid f}(i)\|
      \|\xvec_{\es\mid f}^{\perp} \xvec_{v\mid f}(j) \|} 
\end{equation}
where the second step uses Cauchy Schwarz. Since
\[ \sum_i \|\xvec_{\es\mid f}^{\perp} \xvec_{u\mid f}(i)\| \le
\sqrt{k} \left(\sum_i \|\xvec_{\es\mid f}^{\perp} \xvec_{u\mid
    f}(i)\|^2\right)^{1/2} \le \sqrt{k} \left(\sum_i \| \xvec_{u\mid
    f}(i)\|^2\right)^{1/2} = \sqrt{k} \] the expected value of the
above lower bound \eqref{eq:ali09} for $uv \sim G$ is at least
$\eps_f^2/k$.
\end{proof}

\noindent We now upper bound the lengths of the embedded vectors.
\begin{claim}
  \label{clm:brs-ub}
  $\|\xmat_u(f)\|^2 \le \sum_{i \in [k]} \left\| \xvec_{\es\mid
      f}^{\perp} \xvec_{u\mid f}(i)\right\|^2$. In particular,
  $\mathbb{E}_u \|\xmat_u(f)\|^2 \le {\delta_f}{}$.
\end{claim}
\begin{proof}
  $ \frac{1}{k} \|\xmat_u(f)\|^2 = \mathbb{E}_{ij} \frac{\langle
    \xvec_{\es\mid f}^{\perp} \xvec_{u\mid f}(i),
    \xvec_{\es\mid f}^{\perp} \xvec_{u\mid f}(j) \rangle^2
  }{\|\xvec_{\es\mid f}^{\perp} \xvec_{u\mid f}(i)\| \|\xvec_{\es\mid
      f}^{\perp} \xvec_{u\mid f}(j) \|} \le \left(\mathbb{E}_{i}
    \left\| \xvec_{\es\mid f}^{\perp} \xvec_{u\mid
        f}(i)\right\|\right)^2 \le \mathbb{E}_{i} \left\|
    \xvec_{\es\mid f}^{\perp} \xvec_{u\mid f}(i)\right\|^2.$
\end{proof}

\noindent Now for fixed $f$ we will upper bound the expected value of
$\delta_{f, g}$ over $g \sim \|\xvec_{S\mid f}(g)\|^2$ in terms of the
projection distance of the embedded vectors from the subspace spanned
by $X_v(f)$ for $v \in S$. (Below, $X_S(f)^{\perp}$ denotes the
projection onto the orthogonal complement of $\mathrm{span} \{ X_v(f)
\mid v\in S \}$.)
\begin{claim} \label{lem:brs-pi-to-x} $ \mathbb{E}_{g \sim
    \|\xvec_{S\mid f}(g)\|^2} \left[ \delta_{f,g} \right] \le
  \mathbb{E}_{u\sim G} \left[ \left\| \xmat_S^{\perp}(f)
      \xmat_u(f)\right\|^2 \right]$.
\end{claim}

\begin{proof}
  We know that $ \expcts{g}{ \left\|\xvec_{\es\mid g,f}^{\perp}
      \xvec_{u\mid g, f}(i)\right\|^2 }= \left\| \Pi_{S\mid f}^{\perp}
    \xvec_{u\mid f}(i)\right\|^2$.  Since $\xvec_{\es\mid f}$ is in
  the span of $\Pi_{S\mid f}$
  $\Pi_{S\mid f}^{\perp} \xvec_{u\mid f}(i) = \Pi_{S\mid f}^{\perp}
  \xvec_{\es\mid f}^{\perp} \xvec_{u\mid f}(i)$.  Similarly for any $v
  \in S$ and $j\in [k]$, the vector $\xvec_{\es\mid f}^{\perp}
  \xvec_{v\mid f}(j)$ is in the span of $\Pi_{S\mid f}$.
  By using the same arguments
  from~\Cref{ug-embed-projection-statement}, namely the embedding used
  here preserves linearity, we obtain $ \sum_i \left\|{\Pi}_{S\mid
      f}^{\perp} \xvec_{\es\mid f}^{\perp} \xvec_{u\mid
      f}(i)\right\|^2 \le \left\|\xmat_S^{\perp}(f) \xmat_u(f)
  \right\|^2$.  Taking expectation over $u$ completes the proof.
\end{proof}

\noindent Using the above, we can prove the main claim about the seed
selection procedure, namely that, assuming $\lambda_r$ is close enough
to $1$, the expected variance $\delta_f$ can be reduced by a geometric
factor by conditioning on the assignment to a further $O(r/\eps)$
nodes.
\begin{lemma} \label{lem:brs-001} Given $f\in \labs^{S_0}$, positive
  real $\eps>0$ and positive integer $r$ with $\lambda_{r+1} \ge 1 -
  \frac{\eps^2}{2 k^2}$, if $\epsilon_f \ge \eps$ then there exists a
  set of $O(r k^2/\eps^2)$-columns of $\xmat(f)$, $S$ and $g \in
  \labs^S$ such that $\xvec_{S\mid f}(g)\neq 0$ and:
\begin{equation}
  \delta_{f,g}
  \le 
  \delta_f - \Omega\left(\frac{\eps^2}{k^2}\right).
  \label{eq:lem-brs000}
\end{equation}
\end{lemma}
\begin{proof}
  Let $\rho \triangleq \eps/k$, and $\mu\triangleq \mathbb{E}_u
  \|\xmat_u\|^2$ where for notational convenience we suppress the
  dependence on $f$ and denote $X_u(f)$ by $X_u$.
Observe that \[ \frac{1}{n} \tr\left[ {\xmat}^T {\xmat} A\right] = 
\mathbb{E}_{uv\sim G} \langle \xmat_u, \xmat_v\rangle\ge (\eps_f/k)^2 \ge \rho^2 \]
by~\Cref{clm:brs-lb}. 
This implies $\frac{1}{n} \tr\left[ {\xmat}^T {\xmat} L \right] \le
\mathbb{E}_u \|{\xmat_u}\|^2 - \rho^2 = \mu - \rho^2$.
From \Cref{lem:markov-bound}, %
we know that volume sampling $O(r/\rho^2)$ columns from ${\xmat}$
yields a set $S$ for which:
\begin{align*}\mathbb{E}_u \|\xmat_S^{\perp}{\xmat_u}\|^2
  \le & \left(1+{O(\rho^2)}\right) \frac{(1/n) \tr\left[ {\xmat}^T
      {\xmat} L \right]}{1-\max(1-\lambda_{r+1},0)} \le
  \left(1+O(\rho^2)\right) \frac{\mu - \rho^2}{1 - \frac{\rho^2}{2}}
  \intertext{Since $\rho\le 1$, we have $(1 - \rho^2/2)^{-1} \le
    \left(1+\frac{3}{4}\rho^2\right)$:} \le &
  \left(1+O(\rho^2)\right)(\mu - \rho^2) \left(1 + \frac{3}{4}
    \rho^2\right)
  \le  \left(1+O(\rho^2)\right)\left(\mu - \frac{\rho^2}{4}\right) \\
  \le &  \mu - \Omega\left(\rho^2\right) \\
  \le & \delta_f - \Omega\left(\rho^2\right) \qquad \mbox{(by
    \Cref{clm:brs-ub})} \ .
\end{align*}
By~\Cref{lem:brs-pi-to-x}, $\mathbb{E}_g\left[ \delta_{f,g}\right] \le
\mathbb{E}_u \|\xmat_S^{\perp}\xmat_u\|^2$, which means there exists
$g$ for which $\delta_{f,g} \le \delta_f -
\Omega\left(\frac{\eps^2}{k^2}\right)$.
\end{proof}
We put together everything in the following theorem. 
\begin{theorem} \label{thm:brs-final} For $\ell =
  O\left(\frac{k^2}{\eps^2}\right)$, seed selection procedure will
  output a partial assignment $f_{\ell}$ with $\epsilon_{f_{\ell}} \le
  \eps$.
\end{theorem}
\begin{remark}[Faster Implementation]
  \label{rem:faster-brs}
  By using the faster solver from~\cite{gs12-fast}, we can implement a
  matching algorithm for~\Cref{thm:brs-final} whose running time is
  $2^{\poly(k/\eps) r} n^{\poly(k/\eps)}$.
\end{remark}
\begin{proof}
  Suppose $\epsilon_{f_i} > \eps$ for all $i \le \ell$.  Then
  by~\Cref{lem:brs-001}, for each $i \le \ell$:
  \[
  0\le \delta_{f_{i}} \le \delta_{f_0} - i
  \Omega\left(\frac{\eps^2}{k^2}\right) \le 1 - \Omega\left(\frac{i
      \eps^2}{k^2}\right) \implies \delta_{f_{\ell}} < 0,
  \] which is a contradiction.
\end{proof}

\subsection{Partial Coloring of $3$-Colorable Graphs}
\label{sec:partial-color}
The rounding algorithm is given below:
\begin{enumerate}
\item Let $\xmat_u \gets \sum_{i=1}^3 e_i \otimes \xvec_{\es}^{\perp}
  \xvec_{u}(i)$ (same as in~\cite{ag11}).
\item Use~\Cref{thm:svd} to choose $S$, an $r'$-subset of vectors from
  $\left(\xmat_{u}\right)_{u\in [n]}$.
\item Sample $f\sim \|\xvec_S(f)\|^2$.
\item For each $u\in V$, let $\xint_u \gets j$ if $\|\xvec_{u\mid
    f}(j)\|^2 > \frac12$ for some $j \in \{1,2,3\}$. If no such $j$
  exists, let $\xint_u \gets 0$.
\item Output the partial coloring, $\xint \in \{0,1,2,3\}^V  $.
\end{enumerate}
\begin{theorem} \label{thm:main-ag} Given a $3$-colorable $d$-regular
  graph $G$ on $n$ nodes, positive real $1>\eps>0$ and positive
  integer $r$, suppose its $r^{th}$ largest eigenvalue of normalized
  Laplacian matrix, $\lambda_{n-r}$, satisfies
  \[
  \lambda_{n-r} \le \frac{4-\delta}{3} 
  \] for some positive real $\delta> 0$. Then, for the choice of $r' =
  O(r/\delta\eps)$, by solving $r'$ rounds of~\lh\ relaxation, we can
  find a partial coloring which colors at least $(1-\eps)
  \frac{\delta}{2+\delta} n$ nodes.
\end{theorem}
\begin{remark}[Faster Implementation]
  \label{rem:faster-ag}
  By using the faster solver from~\cite{gs12-fast}, we can implement a
  matching algorithm for~\Cref{thm:main-ag} whose running time is
  $2^{O(r)} n^{O(1)}$.
\end{remark}

We have the following as an immediate corollary
of~\Cref{thm:main-ag}:
\begin{corollary}
Given a $3$-colorable $d$-regular graph $G$,
for any positive integer $r$ with 
$\lambda_{n-r} \le \frac{10}{9}-\Omega(1)$, we can
find a partial coloring on $\frac{n}{4}$ nodes and an independent set of size
at least $\frac{n}{12}$ in time $n^{O(r)}$.
\end{corollary}
Before we begin the proof of~\Cref{thm:main-ag}, we will state some
simple claims. As the method applies for $k$-colorable graphs with
different parameters, below for clarity we first use $k$ for the
number of colors, and then later set $k =3$.
\begin{claim} \label{clm:dist-ag}
For any edge $(u,v)$ of $G$, 
\[
\frac{1}{2} \left\| \xmat_u + \xmat_v\right\|^2 \le 1 - \frac{2}{k}.
\]
In particular, if we use $A$ to denote the normalized adjacency matrix
of $G$, then:
\[
\tr\left[ \xmat^T \xmat (I+A)\right] 
\le n \left(1 - \frac{2}{k}\right).
\]
\end{claim}
\begin{proof}
\begin{align*}
\frac{1}{2}  \left\| \xmat_u + \xmat_v\right\|^2 &= 
	\frac{1}{2}  \sum_{i\in [k]} \Bigl( \|\xvec_{\es}^{\perp} \xvec_{u}(i)\|^2
+ \|\xvec_{\es}^{\perp} \xvec_{v}(i)\|^2 + 2 \langle \xvec_{\es}^{\perp} \xvec_{u}(i), 
	\xvec_{\es}^{\perp} \xvec_{v}(i)\rangle \Bigr)  \\
& =  \frac{1}{2} \sum_{i\in [k]} \biggl( \|\xvec_{u}(i)\|^2-\|\xvec_{u}(i)\|^4
+ \|\xvec_{v}(i)\|^2-\|\xvec_{v}(i)\|^4 \\
&  \quad\quad + 2 \langle \xvec_{u}(i), \xvec_{v}(i)\rangle  
- 2 \langle \xvec_{\es}, \xvec_{u}(i) \rangle \langle 
	\xvec_{\es}, \xvec_{v}(i)\rangle \|\xvec_\es\|^2  \biggr)  
\intertext{Using $\langle x_u(i),x_v(i)\rangle = 0$, we can rewrite this as:}
& =  1 - \frac{1}{2} \sum_{i\in [k]} \left( \|\xvec_{u}(i)\|^4+\|\xvec_{v}(i)\|^4+
 2\|\xvec_{u}(i)\|^2 \|\xvec_{v}(i)\|^2\right) \\
& =  1 - \frac{1}{2} \sum_{i\in [k]} \left( \|\xvec_{u}(i)\|^2+\|\xvec_{v}(i)\|^2\right)^2.
\end{align*}
At this point, observe that $\sum_{i\in [k]} \left(
  \|\xvec_{u}(i)^2+\|\xvec_{v}(i)\|^2\right)^2$ is a convex function
on $\|\xvec_u(i)\|^2$ and $\|\xvec_v(j)\|^2$'s. Since $\sum_i
\|\xvec_u(i)\|^2 = \sum_j \|\xvec_v(j)\|^2 = 1$, it is minimized when
$\|\xvec_u(i)\|^2=\|\xvec_v(j)\|^2=\frac{1}{k}$. Substituting this
into the above expression, we see that:
\[
\frac{1}{2} \left\| \xmat_u + \xmat_v\right\|^2 \le 1 - \frac{k}{2}
\left( \frac{2}{k} \right)^2 = 1 - \frac{2}{k}.
\]
For the final part, observe that:
\[
\tr\left[ \xmat^T \xmat (I+A )\right] = \frac{1}{d} \sum_{\{u,v\}\in
  E(G)} \|\xmat_u + \xmat_v\|^2 \le \frac{2 |E(G)|}{d} \left(1 -
  \frac{2}{k}\right) = n \left(1 - \frac{2}{k}\right)
. \tag*{\qedhere}
\]
\end{proof}
\begin{claim} \label{clm:delta-r-ag} Given a graph $G$ and positive
  integer $r$, for $\lambda_r$ being the $r^{th}$ smallest eigenvalue
  of corresponding normalized graph Laplacian matrix, the following
  holds:
\[
\sum_{j\ge r} \sigma_j(\xmat^T \xmat) \le n \frac{1-2/k}{2-\lambda_r}.
\]
\end{claim}
\begin{proof}
  Follows from using the upper bound from~\Cref{clm:dist-ag} on
  inequality:
  \[
  \sum_{j\ge r} \sigma_j(\xmat^T \xmat) \le \frac{1}{\lambda_{r}}
  \tr\left[ \xmat^T \xmat (I+A)\right] \ . \tag*{\qedhere}
  \]
\end{proof}
\begin{claim}
  Assume $u$ is uncolored. Then:
\[
\sum_i \|\xvec_{\es\mid f}^\perp \xvec_{u\mid f(i)}\|^2 \ge \frac{1}{2}.
\]
\end{claim}
\begin{proof}
  Note that $\|\xvec_{\es\mid f}^\perp \xvec_{u\mid f}(i)\|^2 =
  \|\xvec_{u\mid f}(i)\|^2 (1 - \|\xvec_{u\mid f}(i)\|^2)$. If $u$ is
  uncolored, then $1 - \|\xvec_{u\mid f}(i)\|^2\ge \frac{1}{2}$ for
  all $i\in [k]$\footnote{This follows from the threshold rounding
    algorithm used in \cite{ag11} for coloring, which colors $u$ with
    color $i$ if $\|\xvec_{u\mid f}(i)\|^2 > 1/2$.}, in which case we
  have:
  \[
  \sum_i \|\xvec_{\es\mid f}^\perp \xvec_{u\mid f}(i)\|^2 \ge
  \frac{1}{2} \sum_i \|\xvec_{u\mid f}(i)\|^2 = \frac{1}{2}.\qedhere
  \]
\end{proof}
For a subset $S$ of vertices of $G$, we denote by $X_S^\perp$ the
projection operator onto the orthogonal complement of
$\mathrm{span}\{X_u \mid u \in S\}$.
\begin{lemma}
  If we sample $f:S\to [3]$
  with probability $\|\xvec_S(f)\|^2$:
  \[
  \mathbb{E}_f\left[ \sum_i \|\xvec_{\es\mid f}^\perp \xvec_{u\mid f}(i)\|^2  \right] 
  \le \|\xmat_S^{\perp} \xmat_u\|^2.
  \]
\end{lemma}
\begin{proof}
  Follows from~\Cref{thm:expct-over-f-binary,thm:pi-to-p}.
\end{proof}
\begin{proof}[Proof of~\Cref{thm:main-ag}]
  Let $\delta' = \frac{1}{2} \delta$ and $\eps' = \eps \delta'$.
  By~\Cref{thm:svd}, we know that there exists $r' = O(r/\eps')$
  columns, $S$, such that:
  \[
  \sum_u \|\xmat_S^{\perp} \xmat_u\|^2 \le (1+\eps) \sum_{j\ge r}
  \sigma_j(\xmat^T \xmat) \le n (1+\eps') \frac{1-2/k}{2-\lambda_r}.
  \]
  Using Markov inequality, the fraction of uncolored nodes is bounded
  by:
  \[
  \le 2 n (1+\eps) \frac{1-2/k}{2-\lambda_r} = \frac{2 (1+\eps')}{3
    (2-\lambda_r)} n \quad \text{(for $k=3$).} \] For $\lambda_r \le
  \frac{4}{3} - \frac{2}{3} \delta'$, this expression becomes
  $\frac{1+\eps'}{1+\delta'} n$, which implies
  \[ 
  \mathbb{E}\left[\text{fraction of colored nodes}\right] \ge 1 -
  \frac{1+\eps'}{1+\delta'} = \frac{\delta' - \eps'}{1 + \delta'} =
  \frac{\delta'}{1+\delta'} (1-\eps) = \frac{\delta/2}{1 + \delta/2}
  (1-\eps).%
  \]

  To prove that the coloring output is legal, notice that for any pair
  of adjacent nodes $(u,v)\in E(G)$, both $\|\xvec_{u|f}(i)\|^2$ and
  $\|\xvec_{v|f}(i)\|^2$ cannot be larger than $1/2$ both at the same
  time.
\end{proof}

\section{Preliminaries}
\label{sec:prelim}

\section{Case Study: Minimum Bisection}
\label{sec:case-study}

\section{Analysis of Propagation Rounding}
\label{sec:prop-round}

\section{Choosing A Good Seed Set by Column Selection}
\label{sec:seed-good}

\section{Algorithms Based on Independent Rounding} 
\label{sec:app-independent}
In this section, we will present approximation algorithms which are
based on independent rounding scheme. Unless stated otherwise, our
algorithms will all follow the same outline:
\begin{enumerate}
\item For some matrix $A$, choose seeds $\bestS$ as described
  in~\Cref{thm:bounding-delta-r} with $B \triangleq \diag(A)$.
\item Sample $f:\bestS \to \labs$ with probability $\|\xvec(f)\|^2$. 
\item For each $u\in V$, independently sample a label $j\in \labs$
  from the distribution $(\|\xvec_{u\mid f}(j)\|^2)$.
\end{enumerate}

\subsection{Quadratic Integer Programming}
\label{sec:qip}
\begin{theorem}
  \label{thm:qip}
  Consider a quadratic integer programming problem with objective
  function of the form
  \[
  \xint^T A \xint 
  \] for some $A\in \sympm^{V\times \labs}$ subject to various linear
  constraints. Then by solving $O(r/\eps^2)$-rounds of \lh~relaxation,
  we can obtain a solution $\xint$ such that:
  \begin{itemize}
  \item $\xint^T A \xint \le (1+\eps)
    \frac{\opt}{\lambda_{r}(A,\diag(A))}.$
  \item $\xint$ is obtained by a faithful and independent rounding. 
    In particular, concentration bounds hold for linear constraints.
  \end{itemize}
\end{theorem}
\begin{remark}[Faster Implementation]
  \label{rem:faster-qip}  
  By using the faster solver from~\cite{gs12-fast}, we can implement a
  matching algorithm for~\Cref{thm:qip} whose running time is
  $2^{O(r/\eps^3)} n^{O(1/\eps)}$.
\end{remark}
\begin{proof}
  For fixed $\bestS$, the expected value of the rounding is equal to 
  \[
  \tr(\xmat^T \Pi_{\bestS} \xmat A) + 
  \tr(\xmat^T \Pi_{\bestS}^\perp \xmat \diag(A))
  \] by~\Cref{thm:ind-expct-over-mtx}. If we choose $\bestS$ using
  \Cref{thm:bounding-delta-r} with $B \gets \diag(A)$, then this value
  is upper bounded by:
  \[
  (1+\eps) \frac{\tr(\xmat^T \xmat A)}{\lambda_{r}(A,\diag(A))}. \tag*{\qedhere}
  \]
\end{proof}

\subsection{Maximum Cut and Unique Games}
\label{sec:ug}

\subsection{Independent Set}
\label{sec:is-gen}

\section{Algorithms Based on Threshold Based Rounding}
\label{sec:app-threshold}
In this section, we will present approximation algorithms which use the
threshold based rounding scheme. All problems share the following
structure: Given a capacity graph, $G=(V,C)$, and a \emph{connected} demand
graph, $H=(V,D)$, find  a non-empty $U \subset V$ which minimizes the
total capacity cut subject to various constraints involving the total
demand cut.

Recall that, for $L_H$ and $\Gamma_H$ being the Laplacian and
edge-node incidence matrices of graph $H$, respectively, we have $L_H
= \Gamma_H^T \Gamma_H$. Since $H$ is connected, null space of $L_H$
consists only of constant vector. Hence $\nll(L_H) \subseteq
\nll(L_G)$. Also the width of $\Gamma_H$ is $2$. Therefore matrices
$L_G$ and $L_H$ satisfy all the requirements
in~\Cref{lem:markov-bound}, which we use for seed selection. The
following is the randomized rounding algorithm we will use in this
section:
\begin{enumerate}
\itemsep=0.5ex
\item Choose seeds $\bestS$ as described in~\Cref{lem:markov-bound}
  where $A \triangleq L_G$, $B \triangleq L_H$ and $\Gamma \triangleq
  \Gamma_H$.
\item \label{item:sample-f} Sample $f:\bestS \to \labs$ with
  probability $\|\xvec(f)\|^2$.
\item \label{item:sample-tau}
  Choose a threshold $\tau \in (0,1]$ uniformly at random.
\item For each $u\in V$, set $\xint_u \gets 1$ if $\|\xvec_{u\mid
    f}\|^2 \ge \tau$ and $\xint_u \gets 0$ otherwise.
\end{enumerate}
In all our analysis, we will prove that the procedure will output
correct answer with positive probability. But we can easily
derandomize the above procedure by enumerating over all:
\begin{enumerate}[(i)]
\itemsep=0.5ex
\item $f:\bestS \to \{0,1\}$ in step~(\ref{item:sample-f}),
\item $\tau \in \{ \|\xvec_{u\mid f}\|^2 \mid u \in V\}$ in
  step~(\ref{item:sample-tau}).
\end{enumerate}
After these changes, the running time of rounding is $2^{O(r)}
\poly(n)$, which does not affect the total running time.

\subsection{Sparsest Cut and Variations}
\label{sec:sc}

\subsection{Minimizing Capacity Cut Under Packing Constraints}
\label{sec:qcut}

\section*{Acknowledgments}
We thank Sanjeev Arora for useful comments on an earlier manuscript.
\bibliographystyle{alpha}
\bibliography{references}

\newcommand{\etalchar}[1]{$^{#1}$}
\begin{thebibliography}{AKK{\etalchar{+}}08}

\bibitem[ABS10]{abs}
Sanjeev Arora, Boaz Barak, and David Steurer.
\newblock Subexponential algorithms for {Unique} {Games} and related problems.
\newblock In {\em FOCS}, pages 563--572, 2010.

\bibitem[AG11]{ag11}
Sanjeev Arora and Rong Ge.
\newblock New tools for graph coloring.
\newblock In {\em APPROX-RANDOM}, pages 1--12, 2011.

\bibitem[AKK{\etalchar{+}}08]{akkstv08}
Sanjeev Arora, Subhash Khot, Alexandra Kolla, David Steurer, Madhur Tulsiani,
  and Nisheeth~K. Vishnoi.
\newblock Unique games on expanding constraint graphs are easy.
\newblock In {\em STOC}, pages 21--28, 2008.

\bibitem[AL08]{al08}
Reid Andersen and Kevin~J. Lang.
\newblock An algorithm for improving graph partitions.
\newblock In {\em SODA}, pages 651--660, 2008.

\bibitem[AMS11]{ams11}
Christoph Amb{\"u}hl, Monaldo Mastrolilli, and Ola Svensson.
\newblock Inapproximability results for maximum edge biclique, minimum linear
  arrangement, and sparsest cut.
\newblock {\em SIAM J. Comput.}, 40(2):567--596, 2011.

\bibitem[ARV09]{arv}
Sanjeev Arora, Satish Rao, and Umesh~V. Vazirani.
\newblock Expander flows, geometric embeddings and graph partitioning.
\newblock {\em J. ACM}, 56(2), 2009.

\bibitem[BBH{\etalchar{+}}12]{bbhksz12}
Boaz Barak, Fernando G. S.~L. Brand{\~a}o, Aram~Wettroth Harrow, Jonathan~A.
  Kelner, David Steurer, and Yuan Zhou.
\newblock Hypercontractivity, sum-of-squares proofs, and their applications.
\newblock In {\em STOC}, pages 307--326, 2012.

\bibitem[BDMI11]{bdm11}
Christos Boutsidis, Petros Drineas, and Malik Magdon-Ismail.
\newblock Near optimal column-based matrix reconstruction.
\newblock In {\em FOCS}, pages 305--314, 2011.

\bibitem[BRS11]{brs11}
Boaz Barak, Prasad Raghavendra, and David Steurer.
\newblock Rounding semidefinite programming hierarchies via global correlation.
\newblock In {\em FOCS}, pages 472--481, 2011.

\bibitem[CDK12]{cdk12}
Eden Chlamtac, Michael Dinitz, and Robert Krauthgamer.
\newblock Everywhere-sparse spanners via dense subgraphs.
\newblock In {\em FOCS}, pages 758--767, 2012.

\bibitem[Chl07]{chlamtac}
Eden Chlamtac.
\newblock Approximation algorithms using hierarchies of semidefinite
  programming relaxations.
\newblock In {\em FOCS}, pages 691--701, 2007.

\bibitem[CS08]{cs08}
Eden Chlamtac and Gyanit Singh.
\newblock Improved approximation guarantees through higher levels of {SDP}
  hierarchies.
\newblock In {\em APPROX-RANDOM}, pages 49--62, 2008.

\bibitem[CT11]{ct11}
Eden Chlamtac and Madhur Tulsiani.
\newblock Convex relaxations and integrality gaps.
\newblock In {\em Handbook on Semidefinite, Cone and Polynomial Optimization}.
  Springer, 2011.

\bibitem[DMN13]{dmn13}
Anindya De, Elchanan Mossel, and Joe Neeman.
\newblock Majority is stablest: discrete and sos.
\newblock In {\em STOC}, pages 477--486, 2013.

\bibitem[DR10]{dr10}
Amit Deshpande and Luis Rademacher.
\newblock Efficient volume sampling for row/column subset selection.
\newblock In {\em FOCS}, pages 329--338, 2010.

\bibitem[DV06]{dv06}
Amit Deshpande and Santosh Vempala.
\newblock Adaptive sampling and fast low-rank matrix approximation.
\newblock In {\em APPROX-RANDOM}, pages 292--303, 2006.

\bibitem[Fei02]{feige02}
Uriel Feige.
\newblock Relations between average case complexity and approximation
  complexity.
\newblock In {\em STOC}, pages 534--543, 2002.

\bibitem[GS11]{gs11-qip}
Venkatesan Guruswami and Ali~Kemal Sinop.
\newblock {L}asserre hierarchy, higher eigenvalues, and approximation schemes
  for graph partitioning and quadratic integer programming with {PSD}
  objectives.
\newblock In {\em FOCS}, pages 482--491, 2011.

\bibitem[GS12a]{gs12-fast}
Venkatesan Guruswami and Ali~Kemal Sinop.
\newblock Faster {SDP} hierarchy solvers for local rounding algorithms.
\newblock In {\em FOCS}, pages 197--206, 2012.

\bibitem[GS12b]{gs11-svd}
Venkatesan Guruswami and Ali~Kemal Sinop.
\newblock Optimal column-based low-rank matrix reconstruction.
\newblock In {\em SODA}, pages 1207--1214, 2012.

\bibitem[GS13]{gs11-exp}
Venkatesan Guruswami and Ali~Kemal Sinop.
\newblock Approximating non-uniform sparsest cut via generalized spectra.
\newblock In {\em SODA}, pages 295--305, 2013.

\bibitem[Hal98]{Halldorsson98}
Magn{\'u}s~M. Halld{\'o}rsson.
\newblock Approximations of independent sets in graphs.
\newblock In {\em APPROX}, pages 1--13, 1998.

\bibitem[Hal02]{halperin02}
Eran Halperin.
\newblock Improved approximation algorithms for the vertex cover problem in
  graphs and hypergraphs.
\newblock {\em SIAM J. Comput.}, 31(5):1608--1623, 2002.

\bibitem[H{\aa}s01]{hastad01}
Johan H{\aa}stad.
\newblock Some optimal inapproximability results.
\newblock {\em J. ACM}, 48(4):798--859, 2001.

\bibitem[HJ90]{hj-mat-book}
Roger~A. Horn and Charles~R. Johnson.
\newblock {\em Matrix analysis}.
\newblock Cambridge University Press, 1990.

\bibitem[Kho06]{khot06}
Subhash Khot.
\newblock Ruling out {PTAS} for {G}raph {Min-Bisection}, {Dense k-Subgraph},
  and {Bipartite Clique}.
\newblock {\em SIAM J. Comput.}, 36(4):1025--1071, 2006.

\bibitem[KLPT11]{klpt}
Jonathan~A. Kelner, James~R. Lee, Gregory~N. Price, and Shang-Hua Teng.
\newblock Metric uniformization and spectral bounds for graphs.
\newblock {\em Geometric and Functional Analysis}, 21(5):1117--1143, 2011.

\bibitem[KMN11]{kmn10}
Anna~R. Karlin, Claire Mathieu, and C.~Thach Nguyen.
\newblock Integrality gaps of linear and semi-definite programming relaxations
  for knapsack.
\newblock In {\em IPCO}, volume 6655, pages 301--314. Springer Berlin
  Heidelberg, 2011.

\bibitem[Kol10]{kolla10}
Alexandra Kolla.
\newblock Spectral algorithms for unique games.
\newblock In {\em CCC}, pages 122--130, 2010.

\bibitem[KPS10]{kps-lasserre}
Subhash Khot, Preyas Popat, and Rishi Saket.
\newblock Approximate {L}asserre integrality gap for unique games.
\newblock In {\em APPROX-RANDOM}, pages 298--311, 2010.

\bibitem[KS09]{ks09}
Subhash Khot and Rishi Saket.
\newblock {SDP} integrality gaps with local $\ell_1$-embeddability.
\newblock In {\em FOCS}, pages 565--574, 2009.

\bibitem[KV05]{kv05}
Subhash Khot and Nisheeth~K. Vishnoi.
\newblock The unique games conjecture, integrality gap for cut problems and
  embeddability of negative type metrics into l$_{\mbox{1}}$.
\newblock In {\em FOCS}, pages 53--62, 2005.

\bibitem[Las02]{las02}
Jean~B. Lasserre.
\newblock An explicit equivalent positive semidefinite program for nonlinear
  0-1 programs.
\newblock {\em SIAM J. Optimization}, 12(3):756--769, 2002.

\bibitem[Lau03]{laurent03}
Monique Laurent.
\newblock A comparison of the {S}herali-{A}dams, {L}ov{\'a}sz-{S}chrijver, and
  {L}asserre relaxations for 0-1 programming.
\newblock {\em Math. Oper. Res.}, 28(3):470--496, 2003.

\bibitem[LS91]{ls91}
L{\'a}szl{\'o} Lov{\'a}sz and Alexander Schrijver.
\newblock Cones of matrices and set-functions and 0-1 optimization.
\newblock {\em SIAM J. Optimization}, 1:166--190, 1991.

\bibitem[MM10]{mm10}
Konstantin Makarychev and Yury Makarychev.
\newblock How to play unique games on expanders.
\newblock In {\em WAOA}, pages 190--200, 2010.

\bibitem[Oli10]{oliveira}
Roberto Oliveira.
\newblock The spectrum of random $k$-lifts of large graphs (with possibly large
  $k$).
\newblock {\em Journal of Combinatorics}, 1(3-4):285--306, 2010.

\bibitem[OZ13]{oz13}
Ryan O'Donnell and Yuan Zhou.
\newblock Approximability and proof complexity.
\newblock In {\em SODA}, pages 1537--1556, 2013.

\bibitem[PP93]{pp93}
James~K. Park and Cynthia~A. Phillips.
\newblock Finding minimum-quotient cuts in planar graphs.
\newblock In {\em STOC}, pages 766--775, 1993.

\bibitem[Rag08]{rag08}
Prasad Raghavendra.
\newblock Optimal algorithms and inapproximability results for every {CSP}?
\newblock In {\em STOC}, pages 245--254, 2008.

\bibitem[RS09]{rs09}
Prasad Raghavendra and David Steurer.
\newblock Integrality gaps for strong {SDP} relaxations of {U}nique {G}ames.
\newblock In {\em FOCS}, pages 575--585, 2009.

\bibitem[RS10]{rs10-sse}
Prasad Raghavendra and David Steurer.
\newblock Graph expansion and the unique games conjecture.
\newblock In {\em STOC}, pages 755--764, 2010.

\bibitem[RST12]{rst10}
Prasad Raghavendra, David Steurer, and Madhur Tulsiani.
\newblock Reductions between expansion problems.
\newblock In {\em CCC}, 2012.

\bibitem[RT12]{rt11}
Prasad Raghavendra and Ning Tan.
\newblock Approximating {CSPs} with global cardinality constraints using {SDP}
  hierarchies.
\newblock In {\em SODA}, 2012.

\bibitem[SA90]{sa90}
Hanif~D. Sherali and Warren~P. Adams.
\newblock A hierarchy of relaxations between the continuous and convex hull
  representations for zero-one programming problems.
\newblock {\em SIAM J. Discrete Mathematics}, 3:411--430, 1990.

\bibitem[Sch08]{schoenebeck}
Grant Schoenebeck.
\newblock Linear level {L}asserre lower bounds for certain k-{CSP}s.
\newblock In {\em FOCS}, pages 593--602, 2008.

\bibitem[TSSW00]{tssw00}
Luca Trevisan, Gregory~B. Sorkin, Madhu Sudan, and David~P. Williamson.
\newblock Gadgets, approximation, and linear programming.
\newblock {\em SIAM J. Comput.}, 29(6):2074--2097, 2000.

\bibitem[Tul09]{tulsiani}
Madhur Tulsiani.
\newblock {CSP} gaps and reductions in the {L}asserre hierarchy.
\newblock In {\em STOC}, pages 303--312, 2009.

\end{thebibliography}

\appendix
\section{Analysis of Other Rounding Algorithms Using Column Selection
  Framework}
\label{sec:other-rounding-col-sel}
\subsection{$2$-CSPs}
\label{sec:csp}

\subsection{Partial Coloring of $3$-Colorable Graphs}
\label{sec:partial-color}

\end{document}